\documentclass{ws-ijmpe}
\usepackage{hyperref}
\usepackage{cite}
\usepackage{amsmath}
\usepackage[T1]{fontenc}
\usepackage{mathptmx}
\usepackage{txfonts}
\usepackage{centernot}
\usepackage{tensor}
\usepackage{cancel}
\newcommand{\HCd}{\mathcal{H}}
\newcommand{\HCdtD}{\tilde{\HCd}_{\mathrm{Dyn}}}
\newcommand{\HCdD}{\HCd_{\mathrm{Dyn}}}
\newcommand{\HCdtz}{\tilde{\HCd}_{0}}
\newcommand{\FCd}{\tilde{\mathcal{F}}}
\newcommand{\LCd}{\mathcal{L}}
\newcommand{\dd}{\,\mathrm{d}}
\newcommand{\rmi}{\mathrm{i}}
\newcommand{\onehalf}{{\textstyle\frac{1}{2}}}
\newcommand{\quarter}{{\textstyle\frac{1}{4}}}
\newcommand{\pfrac}[2]{\frac{\partial{#1}}{\partial{#2}}}
\newcommand{\ppfrac}[3]{\frac{\partial^{2}{#1}}{\partial{#2}\partial{#3}}}
\newcommand{\sla}[2]{\left.#1\right|_{#2}}
\newcommand{\xX}[2]{\pfrac{x^{#1}}{X^{#2}}}
\newcommand{\Xx}[2]{\pfrac{X^{#1}}{x^{#2}}}
\begin{document}
\markboth{J.~Struckmeier et al.}{Spacetime coupling of spin-0 and spin-1 particle fields}
%
%
\title{Canonical transformation path to gauge theories of gravity II\\[\medskipamount]
--- Spacetime coupling of spin-0 and spin-1 particle fields ---}
\author{J.~Struckmeier$^{1,2,3}$\footnote{struckmeier@fias.uni-frankfurt.de}\;\,,
J.~Muench$^{1,2}$\footnote{Present address: Faculty of Physics, University of Regensburg, johannes.muench@ur.de}\;\,, P.~Liebrich$^{1,2}$,
M.~Hanauske$^{1,2}$, J.~Kirsch$^{1}$, D.~Vasak$^{1}$, L.~Satarov$^{1,4}$, and H.~Stoecker$^{1,2,3}$}
\address{$^{1}$Frankfurt Institute for Advanced Studies (FIAS), Ruth-Moufang-Strasse~1, 60438 Frankfurt am Main, Germany,\\
$^{2}$Goethe-Universit\"{a}t, Max-von-Laue-Strasse~1, 60438~Frankfurt am Main, Germany\\
$^{3}$GSI Helmholtzzentrum f\"{u}r Schwerionenforschung GmbH, Planckstrasse~1, 64291 Darmstadt, Germany\\
$^{4}$National Research Center ``Kurchatov Institute'', 123183 Moscow, Russia}
\maketitle
\begin{history}
\received{18 November 2018}
\revised{15 January 2019}
\accepted{15 January 2019\\ Published 20 February 2019}
\end{history}
\begin{abstract}
The generic form of spacetime dynamics as a classical gauge field theory has recently been derived,
based on only the action principle and on the Principle of General Relativity.
It was thus shown that Einstein's General Relativity is the special case where (i)
the Hilbert Lagrangian (essentially the Ricci scalar) is supposed to describe the dynamics of
the ``free'' (uncoupled) gravitational field, and (ii) the energy-momentum tensor is that of
scalar fields representing real or complex structureless (spin-$0$) particles.
It followed that all other source fields---such as vector fields representing massive
and non-massive spin-$1$ particles---need careful scrutiny of the appropriate source tensor.
This is the subject of our actual paper: we discuss in detail the coupling of the gravitational
field with (i) a massive complex scalar field, (ii) a massive real vector field, and (iii) a massless vector field.
We show that different couplings emerge for massive and non-massive vector fields.
The \emph{massive} vector field has the \emph{canonical} energy-momentum tensor as the appropriate
source term---which embraces also the energy density furnished by the internal spin.
In this case, the vector fields are shown to generate a torsion of spacetime.
In contrast, the system of a \emph{massless} and charged vector field is associated with the
\emph{metric} (Hilbert) energy-momentum tensor due to its additional $\mathrm{U}(1)$ symmetry.
Moreover, such vector fields do not generate a torsion of spacetime.
The respective sources of gravitation apply for all models
of the dynamics of the ``free'' (uncoupled) gravitational field---which do not follow
from the gauge formalism but must be specified based on separate physical reasoning.
\end{abstract}
\ccode{PACS numbers: 04.50.Kd, 11.15.-q, 47.10.Df\\
DOI: 10.1142/S0218301319500071}
\section{Introduction}
A recent publication\cite{struckmeier17a} (see also~\footnote{Previous attempts to set up a canonical
transformation formalism for a dynamical spacetime in the realm of classical field theory
were published in Refs.\cite{struckmeier13}, \cite{struckmeier15}, and~\cite{struckvasak15}.
Those works are superseded by Ref.\cite{struckmeier17a}.})
presented the formalism of extended canonical transformations in the realm of
covariant Hamiltonian field theory\cite{dedonder30,weyl35,struckmeier08,StrRei12,struckvasak15}.
This enables the description of canonical transformations
of classical fields under general mappings of the spacetime geometry along the well-established
procedures of gauge transformations, dating back to H.~Weyl\cite{weyl18} and W.~Fock\cite{fock26}.
It generalizes the canonical gauge formalism under a static spacetime background,
which was worked out earlier in Refs.~\cite{StrRei12,Koenigstein16,Struckmeier17}.
Here, the gauge formalism means the generalization of a Lorentz-covariance
to a diffeomorphism covariance by introducing appropriate gauge fields.
This corresponds to the transition from special relativity to general relativity.
The gauge fields turned out to be given by the \emph{connection coefficients} of the spacetime
manifold, which contain the complete information on the spacetime curvature and torsion.
With the metric and the connection emerging as independent field variables,
the canonical gauge theory naturally implements the Palatini formulation\cite{palatini19} of General Relativity.
With the canonical conjugate fields of the metric and the connection,
no additional dynamical quantities need to be introduced \emph{ad hoc} as is done in
the Lagrangian formulation of Utiyama\cite{utiyama56}, Kibble\cite{kibble61}, and Sciama\cite{sciama62}.

The canonical gauge formalism has been applied\cite{struckmeier17a} to a generic dynamical
system of scalar and vector fields in curvilinear spacetime including non-metricity and torsion.
The initially Lorentz-invariant integrand of the action functional has been gauged into
a proper (world) scalar density, hence into an invariant under a dynamical spacetime geometry.
It is shown in the following that two cases of invariant action functionals must be distinguished:
\begin{itemize}
 \item If the respective field only couples to the metric, then the \emph{metric} (Hilbert)
 energy-momentum tensor is the appropriate source term for the spacetime dynamics.
 \item If the field couples to both the metric and the connection, then the metric
 energy-momentum tensor is no longer appropriate.
 For the case of the Proca system, we show that the \emph{canonical} energy-momentum
 tensor then provides the correct source of the spacetime dynamics.
\end{itemize}
As common to all gauge theories, the Hamiltonian describing the dynamics of the ``free''
gauge field, i.e., its dynamics in the absence of any coupling, must be inserted ``by hand''.
In the Hamiltonian representation of $\mathrm{U}(1)$ and Yang-Mills gauge theories\cite{StrRei12},
the dynamics of the ``free'' gauge field is described by a gauge-invariant
term which is quadratic in the canonical momenta of the gauge fields.
For the actual case of the free gravitational field, a Hamiltonian has been
proposed\cite{struckmeier17a}, which is---at most---quadratic in the canonical
momentum tensor of the gauge field---in analogy to the field theories mentioned above.
This quadratic momentum term is actually \emph{required} in order for the correlation of the
momenta to the spacetime derivatives of the gauge field to be uniquely determined.
It can be regarded as a correction term to the Einstein tensor.

The actual paper develops the physical implications of the covariant canonical
gauge theory of gravitation, presented in our previous paper\cite{struckmeier17a}.
In Sect.~\ref{sec:feqs}, a brief review of its results is outlined.
We proceed with the restriction of the obtained field equations to
\emph{metric compatibility}, hence to a covariantly constant metric.
By means of two examples, namely the Klein-Gordon and the Proca system, we demonstrate
the consequences of the different couplings of scalar and vector fields with spacetime.
The gauge formalism will be shown to uniquely determine the appropriate energy-momentum
tensor which acts as the source term for the field equation of gravitation.
This is a critical issue of the theory: as the energy-momentum tensor enters \emph{directly}
into the field equations, it is not admissible to add a zero-divergence term to the source term---as
is done, for instance, by applying the Belinfante-Ro\-senfeld symmetrization
method\cite{belinfante39,rosenfeld40} of the energy-momen\-tum tensor.
We will show that for general vector particle fields, the \emph{canonical} energy-momentum tensor provides
the appropriate source term for the dynamics of the spacetime geometry---even if we neglect torsion.
In contrast to Einstein's general relativity, the spin-$1$ field acts also as a source for torsion of spacetime.
This coupling of spin and torsion is described by a Poisson-type equation,
which is why the torsion can \emph{propagate} along with the gravitational wave.
Hence, for vector particle fields, we encounter from gauge theory a different coupling
of the source field and spacetime as from Einstein's theory.
It is thus also critical for certain astrophysical considerations.
Hence, compact astrophysical objects, like neutron stars and binary neutron star mergers
must be reinvestigated with the appropriate canonical energy-momentum terms for the vector
repulsion effective field theory (EFT).
A similar conclusion does also hold for fermions, both for protons and electrons, as well as
for neutrinos, both in white dwarfs, neutron stars and in ``ultra high energy cosmic ray'' (UHECR) events.
This will be published by the authors in a forthcoming paper.

We then discuss in Sect.~\ref{sec:u1dyn} a system of a
complex scalar field which couples minimally to a massless vector field.
Such a system is commonly referred to as a $\mathrm{U}(1)$ gauge theory.
We work out the canonical gauge theory of gravity for such a system with $\mathrm{U}(1)$ symmetry.
It is shown that the resulting system with $\mathrm{U}(1)\times\mathrm{Diff}(M)$ symmetry
has the metric (Hilbert) energy-momentum tensor as the source of gravity---just as the in case of scalar fields.
Moreover, it turns out that the vector (Maxwell) field does not act as a source for a torsion of spacetime in this case.
This also applies for the gauge vector bosons of $\mathrm{SU}(N)$ (Yang-Mills) gauge theories\cite{strgrei19}.

Finally, the implications of the modified theory on the description of neutron star and black hole mergers
are discussed in the Conclusions (Sect.~\ref{sec:conclusions}).
\section{The field equations of canonical gauge theory of gravity\label{sec:feqs}}
\subsection{Action functionals invariant under the Diff(M) gauge group\label{sec:Diff(M)}}
The dynamics of systems of complex scalar fields $\phi$
and real vector fields $a_{\nu}$ in a Minkowski spacetime background---with their respective conjugates,
$\tilde{\pi}^{\mu}$ and $\tilde{p}^{\nu\mu}$---are described in the covariant Hamiltonian formalism
by a Hamiltonian scalar density $\HCdtz\big(\phi,\phi^{*}\!,\tilde{\pi}^{\mu},\tilde{\pi}^{*\,\mu}\!,a_{\nu},\tilde{p}^{\nu\mu}\!,g_{\mu\nu}\big)$.
The particular canonical field equations follow from the variation of the action
\begin{align}
S&=\int_{R}\left(\pfrac{\phi^{*}}{x^{\,\beta}}\tilde{\pi}^{\,\beta}+\tilde{\pi}^{*\,\beta}\pfrac{\phi}{x^{\,\beta}}+
\tilde{p}^{\,\alpha\beta}\pfrac{a_{\alpha}}{x^{\,\beta}}+\tilde{k}^{\,\alpha\lambda\beta}\pfrac{g_{\alpha\lambda}}{x^{\,\beta}}
+\tilde{q}\indices{_{\eta}^{\alpha\xi\beta}}\pfrac{\gamma\indices{^{\eta}_{\alpha\xi}}}{x^{\,\beta}}-\HCdtz\right)\dd^{4}x.
\label{eq:action-integral2}
\end{align}
Herein, $\tilde{\pi}^{\,\beta}=\pi^{\,\beta}\sqrt{-g}$ and
$\tilde{p}^{\,\alpha\beta}=p^{\,\alpha\beta}\sqrt{-g}$ denote tensor densities
formed from the (absolute) canonical momentum tensors, $\pi^{\,\beta}$ and $p^{\,\alpha\beta}$,
and the determinant, $g=\det(g_{\mu\nu})$, of the system's covariant metric.
Accordingly, $\HCdtz=\HCd_{0}\sqrt{-g}$ denotes the scalar Hamiltonian density
constructed from the ordinary scalar Ham\-iltonian $\HCd_{0}$.
This ensures that the action functional is invariant under arbitrary coordinate transformations.

A closed description of the coupled dynamics of fields and spacetime geometry has been
derived in Ref.\cite{struckmeier17a}, where the gauge formalism yields, on the basis
of Eq.~(\ref{eq:action-integral2}), the amended covariant action functional
\begin{align}
S=\int_{R}\left(\pfrac{\phi^{*}}{x^{\,\beta}}\tilde{\pi}^{\,\beta}+\tilde{\pi}^{*\,\beta}\pfrac{\phi}{x^{\,\beta}}+
\tilde{p}^{\,\alpha\beta}\,a_{\alpha;\beta}+\tilde{k}^{\,\alpha\lambda\beta}\,g_{\alpha\lambda;\beta}
-\onehalf\tilde{q}\indices{_{\eta}^{\alpha\xi\beta}}R\indices{^{\eta}_{\alpha\xi\beta}}-\HCdtD-\HCdtz\right)\dd^{4}x.
\label{eq:action-integral4}
\end{align}
Here $R\indices{^{\eta}_{\alpha\xi\beta}}$ denotes the Riemann-Christoffel curvature tensor
\begin{equation}\label{eq:riemann-tensor}
R\indices{^{\eta}_{\alpha\xi\beta}}=\pfrac{\gamma\indices{^{\eta}_{\alpha\beta}}}{x^{\xi}}-
\pfrac{\gamma\indices{^{\eta}_{\alpha\xi}}}{x^{\beta}}+
\gamma\indices{^{\tau}_{\alpha\beta}}\gamma\indices{^{\eta}_{\tau\xi}}-
\gamma\indices{^{\tau}_{\alpha\xi}}\gamma\indices{^{\eta}_{\tau\beta}},
\end{equation}
which is an abbreviation of this particular combination of
the gauge fields $\gamma\indices{^{\,\xi}_{\alpha\beta}}$ and their spacetime derivatives.
The tensor densities $\tilde{k}^{\,\alpha\lambda\beta}$ and $\tilde{q}\indices{_{\eta}^{\alpha\xi\beta}}$
denote the canonical momenta of the metric $g_{\alpha\lambda}$ and of the connection coefficient
$\gamma\indices{^{\eta}_{\alpha\xi}}$, respectively.
From Eq.~(\ref{eq:action-integral4}) we conclude that $\tilde{k}^{\,\alpha\lambda\beta}$ must be
symmetric in $\alpha$ and $\lambda$ while $\tilde{q}\indices{_{\eta}^{\alpha\xi\beta}}$ must be
skew-symmetric in $\xi$ and $\beta$ as only those parts contribute to the action.
The canonical gauge procedure indeed reproduces the usual minimal coupling substitution,
which converts the partial derivatives of tensors into covariant derivatives---with one important exception:
here, the place of the (nonexistent) covariant derivative of the connection coefficient is taken over by the Riemann tensor.
As common to all gauge theories, a ``dynamics Hamiltonian'',
$\HCdD=\HCdD\big(\,\tilde{k}^{\,\alpha\lambda\beta},\tilde{q}\indices{_{\eta}^{\alpha\xi\beta}},g_{\mu\nu}\big)$,
describing the ``free kinetics'' of the metric and the gauge fields, must be added to the action
integrand ``by hand'' in order to render the gauge fields dynamic quantities.

Furthermore, the \emph{torsion tensor}
\begin{equation}\label{eq:torsion-tensor}
s\indices{^{\,\xi}_{\beta\alpha}}=\onehalf\left(\gamma\indices{^{\,\xi}_{\beta\alpha}}-
\gamma\indices{^{\,\xi}_{\alpha\beta}}\right)\equiv\gamma\indices{^{\,\xi}_{[\beta\alpha]}}
\end{equation}
needs to be considered if we admit non-symmetric connection coefficients.
Schr\"odinger\cite{schroedinger50}, Sciama\cite{sciama62}, and von~der~Heyde \cite{vonderheyde75} showed that
the equivalence principle holds even for spacetime geometries \emph{with} torsion---in contrast to many statements in the literature.

The complete set of field equations for the scalar field $\phi$ and vector field
$a_{\nu}$, coupled to a dynamical spacetime geometry described by the metric $g_{\nu\mu}$,
the connection coefficients $\gamma\indices{^{\,\xi}_{\alpha\beta}}$, and their respective conjugates,
$\tilde{k}^{\lambda\alpha\beta}$ and $\tilde{q}\indices{_{\xi}^{\alpha\beta\eta}}$, then write\cite{struckmeier17a}
\begin{align}
\phi^{*}_{;\mu}&=\pfrac{\HCdtz}{\tilde{\pi}^{\mu}},&
\tilde{\pi}\indices{^{\,\beta}_{;\,\beta}}&=-\pfrac{\HCdtz}{\phi^{*}}+
2\,\tilde{\pi}^{\,\beta}s\indices{^{\,\alpha}_{\beta\alpha}}\label{eq:feq-scalfield}\\
\phi_{;\mu}&=\pfrac{\HCdtz}{\tilde{\pi}^{*\,\mu}},&
\tilde{\pi}\indices{^{*\,\beta}_{;\,\beta}}&=-\pfrac{\HCdtz}{\phi}+
2\,\tilde{\pi}^{*\,\beta}s\indices{^{\,\alpha}_{\beta\alpha}}\label{eq:feq-scalfield2}\\
a_{\nu;\,\mu}&=\pfrac{\HCdtz}{\tilde{p}^{\nu\mu}},&
\tilde{p}\indices{^{\nu\beta}_{;\,\beta}}&=-\pfrac{\HCdtz}{a_{\nu}}+
2\tilde{p}^{\,\nu\beta}s\indices{^{\,\alpha}_{\beta\alpha}}\label{eq:feq-vecfield}\\
g_{\xi\lambda;\,\mu}&=\pfrac{\HCdtD}{\tilde{k}\indices{^{\xi\lambda\mu}}},&
\tilde{k}\indices{^{\,\xi\lambda\beta}_{;\,\beta}}&=-\pfrac{\HCdtz}{g_{\xi\lambda}}-
\pfrac{\HCdtD}{g_{\xi\lambda}}+2\tilde{k}^{\,\xi\lambda\beta}
s\indices{^{\,\alpha}_{\beta\alpha}}\label{eq:em-tensor-den}
\end{align}
and
\begin{align}
\tilde{q}\indices{_{\eta}^{\xi\lambda\beta}_{;\,\beta}}&=
-a_{\eta}\tilde{p}^{\,\xi\lambda}-2g_{\eta\beta}\tilde{k}^{\,\beta\xi\lambda}+
\tilde{q}\indices{_{\eta}^{\xi\beta\alpha}}s\indices{^{\,\lambda}_{\beta\alpha}}+
2\tilde{q}\indices{_{\eta}^{\xi\lambda\beta}}s\indices{^{\alpha}_{\beta\alpha}}\label{eq:feq-conn-coeff3}\\
\onehalf R\indices{^{\eta}_{\xi\lambda\mu}}&=-\pfrac{\HCdtD}
{\tilde{q}\indices{_{\eta}^{\xi\lambda\mu}}}.\label{eq:feqs-Hdyn}
\end{align}
Equation~(\ref{eq:feqs-Hdyn}) links the canonical momentum
$\tilde{q}\indices{_{\eta}^{\alpha\xi\beta}}$ to the Riemann tensor $R\indices{^{\eta}_{\alpha\xi\beta}}$.
As $\HCdtD$ is a scalar (to be exact: a proper \emph{world scalar density}),
we conclude from Eq.~(\ref{eq:feqs-Hdyn}) that the canonical momentum $\tilde{q}\indices{_{\eta}^{\xi\lambda\mu}}$
has the same symmetry properties as the Riemann tensor $R\indices{^{\eta}_{\xi\lambda\mu}}$.
The vacuum field equations, hence a subset of equations~(\ref{eq:feq-scalfield}) to~(\ref{eq:feqs-Hdyn}),
was derived earlier by Nester\cite{nester08} in the language of modern differential geometry by requiring
a diffeomorphism-invariant action.
\subsection{Energy-momentum balance equation}
Inserting Eqs.~(\ref{eq:feq-vecfield}), (\ref{eq:em-tensor-den}), and (\ref{eq:feqs-Hdyn}) into
Eq.~(\ref{eq:feq-conn-coeff3}), it can be covariantly differentiated with respect
to $x^{\,\lambda}$ to get the consistency condition\cite{struckmeier17a}, which actually
represents an energy-momentum balance equation
\begin{align}
2\tilde{k}^{\lambda\alpha\beta}\!\pfrac{\HCdtD}{\tilde{k}^{\lambda\xi\beta}}-2\pfrac{\HCdtD}{g_{\alpha\beta}}g_{\xi\beta}+
\tilde{q}\indices{_{\tau}^{\alpha\beta\lambda}}\!\pfrac{\HCdtD}{\tilde{q}\indices{_{\tau}^{\xi\beta\lambda}}}-
\pfrac{\HCdtD}{\tilde{q}\indices{_{\alpha}^{\tau\beta\lambda}}}\tilde{q}\indices{_{\xi}^{\tau\beta\lambda}}
\!=\pfrac{\HCdtz}{a_{\alpha}}a_{\xi}-\tilde{p}^{\,\alpha\beta}\!\pfrac{\HCdtz}{\tilde{p}^{\,\xi\beta}}+
2\pfrac{\HCdtz}{g_{\alpha\beta}}g_{\xi\beta}.
\label{eq:consistency}
\end{align}
The proof of this equation was originally worked out in the partial derivative representation in\cite{struckmeier17a}.
It is proved directly as a tensor equation in~\ref{app:ricci-formula}.

The terms on the right-hand side of Eq.~(\ref{eq:consistency}) are determined by the
Hamiltonian $\HCdtz$ of the given system of scalar fields $\phi$ and vector fields $a_{\mu}$.
It will be shown that these terms constitute the \emph{canonical} energy-momentum tensor
of the system described by $\HCdtz$.

The terms on the left-hand side emerge from the Hamiltonian $\HCdtD$, which
describes the ``kinetics'' of spacetime.
Hence, the consistency equation~(\ref{eq:consistency}) describes the coupling of the spacetime dynamics
to the dynamics of the source fields as a rank two tensor equation.

The consistency condition~(\ref{eq:consistency})---which follows from the set
of field equations---represents a generic equation of general relativity which holds for
any given system of scalar and vector fields, as described by $\HCdtz$, and for any particular
model for the dynamics of the ``free'' gravitational fields, as described by $\HCdtD$.
The entire set of ten field equations~(\ref{eq:feq-scalfield}) to~(\ref{eq:feqs-Hdyn})
is closed and can be integrated to yield the combined dynamics of fields and spacetime
geometry only after $\HCdtz$ and $\HCdtD$ have been specified.
Note that Eq.~(\ref{eq:consistency}) is \emph{not} restricted to metric compatibility,
hence the covariant derivative of the metric may be nonzero ($g_{\xi\lambda;\mu}\not\equiv0$).
Equation~(\ref{eq:consistency}) also applies for spacetimes with torsion ($s\indices{^{\,\xi}_{\beta\alpha}}\not\equiv0$).

In the proof of Eq.~(\ref{eq:consistency}), the Riemann tensor is not assumed to be skew-symmetric
in its \emph{first} index pair~\footnote{see Misner~et~al.\cite{misner}, p.~324, for a discussion of this issue.}.
This extra symmetry of the Riemann tensor does not follow directly from its definition in Eq.~(\ref{eq:riemann-tensor}).
Without this assumption, an additional term, proportional to the weight factor $w$, appears in the Ricci
formula~(\ref{eq:ricci-formula}) for the commutator of the second covariant derivatives of relative
tensors---which here happens to simplify the proof.
\subsection{Metric compatibility}
If $\HCdtD$ is defined to not depend on the
conjugate of the metric, $\tilde{k}\indices{^{\,\alpha\lambda\beta}}$, then
Eq.~(\ref{eq:em-tensor-den}) establishes the \emph{metric compatibility condition}
\begin{equation}\label{eq:metricity}
g_{\alpha\lambda;\,\beta}=\pfrac{\HCdtD}{\tilde{k}\indices{^{\,\alpha\lambda\beta}}}\equiv Q_{\alpha\lambda\beta}=0,
\end{equation}
wherein $Q_{\lambda\xi\beta}$ denotes the \emph{nonmetricity tensor}.
This reflects a general feature of the canonical formalism: if a Hamiltonian
does not depend on a dynamical variable, then the conjugate variable is conserved.
The restriction to a covariantly constant metric greatly simplifies the subsequent field equations.
If the system Hamiltonian $\HCd_{0}$ does not depend on the metric's conjugate momentum
$\tilde{k}\indices{^{\,\lambda\xi\beta}}$---which corresponds to a system Lagrangian $\LCd_{0}$
that does not depend on the covariant derivative of the metric---the last term on the right-hand side of
Eq.~(\ref{eq:consistency}) represents Hilbert's metric energy-momentum tensor density\cite{hilbert15}
\begin{equation}\label{eq:metric-em}
\tilde{T}^{\alpha\beta}=2\pfrac{\HCdtz}{g_{\alpha\beta}}.
\end{equation}
With a covariantly constant metric, i.e.\ with metric compatibility,
the index $\eta$ in the field equation~(\ref{eq:feq-conn-coeff3}) can simply be raised.
Furthermore, the covariantly constant factor $\sqrt{-g}$ (denoted by the tilde) can be eliminated to yield
\begin{equation}
q\indices{^{\eta\xi\lambda\beta}_{;\,\beta}}=
-a^{\eta}p^{\,\xi\lambda}-2k^{\eta\xi\lambda}+
q\indices{^{\eta\xi\beta\alpha}}s\indices{^{\,\lambda}_{\beta\alpha}}+
2q\indices{^{\eta\xi\lambda\beta}}s\indices{^{\alpha}_{\beta\alpha}}.
\label{eq:feq-conn-coeff4}
\end{equation}
As noticed above, $q\indices{^{\eta\xi\lambda\beta}}$ must be skew-symmetric
in $\eta$ and $\xi$ in order to obey the same symmetries as the Riemann tensor.
In contrast, $k^{\eta\xi\lambda}$ is symmetric, whereas $a^{\eta}p^{\,\xi\lambda}$ has no symmetry in this index pair.
Hence, Eq.~(\ref{eq:feq-conn-coeff4}) actually splits into two equations,
namely the symmetric and the skew-symmetric portion of Eq.~(\ref{eq:feq-conn-coeff4}) in $\eta$ and $\xi$:
\begin{align}
q\indices{^{\eta\xi\lambda\beta}_{;\,\beta}}&=-\onehalf\left(a^{\eta}p^{\,\xi\lambda}-a^{\xi}p^{\,\eta\lambda}\right)+
q\indices{^{\eta\xi\beta\alpha}}s\indices{^{\,\lambda}_{\beta\alpha}}+
2q\indices{^{\eta\xi\lambda\beta}}s\indices{^{\alpha}_{\beta\alpha}}\label{eq:feq-q}\\
2k^{\eta\xi\lambda}&=-\onehalf\left(a^{\eta}p^{\,\xi\lambda}+a^{\xi}p^{\,\eta\lambda}\right)\label{eq:feq-k}.
\end{align}
Either equation will be discussed separately in the following sections.
\subsection{Canonical energy-momentum tensor as the source of gravity}
As $\HCdtD$ is assumed not to depend on $\tilde{k}^{\,\xi\lambda\mu}$ to achieve
metric compatibility ($g_{\xi\lambda;\,\mu}=0$), field equations do not provide
an equation relating $\tilde{k}^{\,\xi\lambda\mu}$ to the derivative
$\partial g_{\xi\lambda}/\partial x^{\mu}$ of the metric.
Nevertheless, the covariant divergence of $\tilde{k}^{\,\xi\lambda\beta}$ from
Eq.~(\ref{eq:em-tensor-den}) and the canonical equations~(\ref{eq:feq-vecfield})
can be inserted into the covariant $\lambda$-derivative of Eq.~(\ref{eq:feq-k}) to yield
\begin{align}
-\frac{2}{\sqrt{-g}}\pfrac{\HCdtD}{g_{\mu\nu}}=
\frac{2}{\sqrt{-g}}\pfrac{\HCdtz}{g_{\mu\nu}}&-\frac{1}{2}\left(g^{\nu\alpha}p^{\mu\beta}+
g^{\mu\alpha}p^{\nu\beta}\right)\pfrac{\HCd_{0}}{p^{\alpha\beta}}
+\frac{1}{2}\left(a^{\mu}\pfrac{\HCd_{0}}{a_{\nu}}+a^{\nu}\pfrac{\HCd_{0}}{a_{\mu}}\right).\label{eq:consistency1}
\end{align}
For the cases considered here, the right-hand side of Eq.~(\ref{eq:consistency1})
sums up to the symmetrized \emph{canonical} energy-momentum tensor $\theta^{\,\mu\nu}$
\begin{equation}\label{eq:cons-rhs}
\theta^{\,\mu\nu}=T^{\mu\nu}-g^{\mu\alpha}p^{\nu\beta}\pfrac{\HCd_{0}}{p^{\alpha\beta}}+a^{\mu}\pfrac{\HCd_{0}}{a_{\nu}}.
\end{equation}
The \emph{canonical} energy-momentum tensor $\theta^{\,\mu\nu}$
follows for a system of complex scalar fields and real vector fields
in the covariant Hamiltonian formalism from the general prescription
\begin{align}
\theta^{\,\mu\nu}&=\pi^{\,\mu}\pfrac{\HCd_{0}}{\pi_{\nu}}+\pi^{*\,\mu}\pfrac{\HCd_{0}}{\pi^{*}_{\nu}}+
g^{\nu\beta}p^{\,\alpha\mu}\pfrac{\HCd_{0}}{p^{\alpha\beta}}+g^{\mu\nu}\left(\HCd_{0}-
\pi^{\alpha}\pfrac{\HCd_{0}}{\pi^{\alpha}}-\pi^{*\,\alpha}\pfrac{\HCd_{0}}{\pi^{*\,\alpha}}-
p^{\,\alpha\beta}\pfrac{\HCd_{0}}{p^{\alpha\beta}}\right).
\label{eq:u1-can-em0}
\end{align}
Equation~(\ref{eq:cons-rhs}) with $\theta^{\,\mu\nu}$ from
Eq.~(\ref{eq:u1-can-em0}) is verified for a Proca system in Eq.~(\ref{eq:proca-emt}).
Hence, our canonical gauge theory of gravity shows that it is exactly the canonical energy-momentum tensor
which constitutes the proper source term of gravity.
This does not apply to a system of scalar and massless vector fields with additional
$\mathrm{U}(1)$ symmetry, as will be shown in Sect.~\ref{sec:u1dyn}.
Both energy-momentum tensors, the metric and the canonical one, differ by terms which are related to the vector field.
Hence, the tensors coincide in the case of systems of pure scalar fields, as is verified in Sect.~\ref{sec:KG-system}.
As the energy-momentum tensor enters \emph{directly} into the field equation of gravity,
it is \emph{not} allowed to replace the canonical energy-momentum tensor by the
metric one if the system comprises a massive or non-massive vector field.

Hence, inserting Eq.~(\ref{eq:cons-rhs}) into Eq.~(\ref{eq:consistency1}) yields
\begin{equation}
\pfrac{\HCdtD}{g_{\mu\nu}}=-\quarter\left(\theta^{\,\mu\nu}+\theta^{\,\nu\mu}\right).
\label{eq:feq-k2}
\end{equation}
Remarkably, it is exactly the symmetric portion of the canonical energy-momentum tensor that acts
as a source for the symmetric tensor on the left-hand side.
Equation~(\ref{eq:feq-k2}) can be regarded as a generic Einstein-type equation which
holds for \emph{any} model of the free (uncoupled) gravitational field, as described by $\HCdtD$.
It also restates the \emph{zero-energy principle}\cite{feynman62},
hence the hypothesis that the average density of matter in the universe has exactly the
\emph{critical} value such that the total energy of the universe is zero.
Actual data suggests that this might indeed be the case\cite{planck15}.

A particular choice of $\HCdtD$ with at most quadratic terms in the canonical
momentum $\tilde{q}\indices{^{\,\xi\tau\beta\lambda}}$ was presented in Ref.\cite{struckmeier17a}.
A more general case is discussed in Ref.~\cite{benisty18c}.

For a scalar field $\phi$ and for restricting the resulting field equation to \emph{linear}
terms in the Riemann tensor, Eq.~(\ref{eq:feq-k2}) yields the Einstein equation.
However, in all other cases a different theory of gravity emerges.

The correlation of \emph{metric} and \emph{canonical} energy-momentum tensors
was discussed by Belinfante\cite{belinfante39} and Rosenfeld\cite{rosenfeld40}.
Their symmetrization prescription has the geometrical meaning to subtract the spin-related components from the canonical
energy-momentum tensor $\theta^{\mu\nu}$---and thereby eliminates the spin-related interactions from the field equations.
\subsection{Spin tensor as the source of spacetime torsion\label{sec:feqs-mc}}
The \emph{skew-symmetric} part of the product $a^{\eta}p^{\xi\lambda}$ in $\eta$ and $\xi$ defines the \emph{canonical spin tensor}
\begin{equation}\label{eq:spintensor}
\tau^{\eta\xi\lambda}=\onehalf\left(a^{\eta}p^{\xi\lambda}-a^{\xi}p^{\eta\lambda}\right),
\end{equation}
which quantifies the \emph{intrinsic} angular momentum (i.e., the spin) density of the vector field $a^{\mu}$\cite{hehl76b,greiner96}.
The tensor $\tau^{\eta\xi\lambda}$ acts as the source term as can be seen by re-writing
Eq.~(\ref{eq:feq-q}) in the form of a Poisson-type equation
\begin{equation}\label{eq:feq-q2}
q\indices{^{\eta\xi\lambda\beta}_{;\,\beta}}-q\indices{^{\eta\xi\alpha\beta}}s\indices{^{\,\lambda}_{\alpha\beta}}-
2q\indices{^{\eta\xi\lambda\alpha}}s\indices{^{\beta}_{\alpha\beta}}=-\tau^{\eta\xi\lambda}.
\end{equation}
Depending on the particular model $\HCdtD$ for the free gravitational field, the canonical momentum $q^{\eta\xi\lambda\beta}$
is correlated in a specific way with the Riemann curvature tensor according to Eq.~(\ref{eq:feqs-Hdyn}).
Hence, Eq.~(\ref{eq:feq-q2}) shows that the spin $\tau^{\eta\xi\lambda}$ may act as a specific source
for the dynamics of a skew-symmetric part of the connection $\gamma\indices{^{\,\lambda}_{\beta\alpha}}$.
As that torsion constitutes an intrinsic property of the Riemann tensor, it \emph{propagates} with gravitational waves.

The second covariant derivative of Eq.~(\ref{eq:feq-q}) yields immediately the skew-symmetric part of the consistency equation
\begin{equation}
R\indices{^{\,\mu}_{\alpha\beta\eta}}\,q\indices{^{\nu\alpha\beta\eta}}-
q\indices{^{\,\mu\alpha\beta\eta}}R\indices{^{\nu}_{\alpha\beta\eta}}=\theta^{\,\mu\nu}-\theta^{\,\nu\mu}.
\label{eq:feq-k3}
\end{equation}
\subsection{Example~1: Complex Klein-Gordon system\label{sec:KG-system}}
The Klein-Gordon Hamiltonian $\HCdtz\big(\phi,\phi^{*},\tilde{\pi}^{\mu},\tilde{\pi}^{*\,\mu},g_{\mu\nu}\big)$
for a system of complex fields in a dynamic spacetime is given by
\begin{equation}\label{eq:kg-ham0}
\HCdtz=\tilde{\pi}^{*\,\alpha}\tilde{\pi}^{\,\beta}\,g_{\alpha\beta}\frac{1}{\sqrt{-g}}+m^{2}\,\phi^{*}\phi\sqrt{-g},
\end{equation}
The set of canonical equations following from~(\ref{eq:kg-ham0}) are
\begin{align*}
\pfrac{\phi}{x^{\nu}}&=\hphantom{-}\pfrac{\HCdtz}{\tilde{\pi}^{*\,\nu}}=
g_{\nu\beta}\frac{\tilde{\pi}^{\beta}}{\sqrt{-g}}=\pi_{\nu}\\
\pfrac{\phi^{*}}{x^{\nu}}&=\hphantom{-}\pfrac{\HCdtz}{\tilde{\pi}^{\nu}}=
g_{\alpha\nu}\frac{\tilde{\pi}^{*\,\alpha}}{\sqrt{-g}}=\pi^{*}_{\nu}\\
\pfrac{\tilde{\pi}^{\alpha}}{x^{\alpha}}&=-\pfrac{\HCdtz}{\phi^{*}}=-m^{2}\phi\sqrt{-g}\\
\pfrac{\tilde{\pi}^{*\,\alpha}}{x^{\alpha}}&=-\pfrac{\HCdtz}{\phi}=-m^{2}\,\phi^{*}\sqrt{-g}.
\end{align*}
The canonical momenta can be eliminated by inserting the momenta
into the equations for the divergence of the momenta
\begin{align*}
\pfrac{\tilde{\pi}^{\alpha}}{x^{\alpha}}=\pfrac{}{x^{\alpha}}
\left(g^{\alpha\beta}\pfrac{\phi}{x^{\beta}}\sqrt{-g}\right)=-m^{2}\phi\sqrt{-g},
\end{align*}
hence
\begin{align*}
g^{\alpha\beta}\ppfrac{\phi}{x^{\alpha}}{x^{\beta}}+\pfrac{\phi}{x^{\beta}}\frac{1}{\sqrt{-g}}
\pfrac{}{x^{\alpha}}\left(g^{\alpha\beta}\sqrt{-g}\right)+m^{2}\phi=0.
\end{align*}
The second term can be expressed in terms of the connection as
\begin{align*}
g^{\alpha\beta}\left(\ppfrac{\phi}{x^{\alpha}}{x^{\beta}}-
\pfrac{\phi}{x^{\xi}}\gamma\indices{^{\xi}_{\alpha\beta}}\right)-
2\pfrac{\phi}{x^{\beta}}g^{\beta\xi}s\indices{^{\alpha}_{\xi\alpha}}+m^{2}\phi=0,
\end{align*}
with $s\indices{^{\alpha}_{\xi\alpha}}$ the contracted torsion tensor, referred to as the \emph{torsion vector}.
The sum in parentheses is the covariant $x^{\beta}$-derivative of the covector $\partial\phi/\partial x^{\alpha}$,
which finally yields the tensor equation
\begin{equation}
\label{eq:KG-feq}
g^{\alpha\beta}\left[{\left(\pfrac{\phi}{x^{\beta}}\right)}_{;\alpha}-
2s\indices{^{\xi}_{\alpha\xi}}\pfrac{\phi}{x^{\beta}}\right]+m^{2}\phi=0.
\end{equation}
The term related to the torsion vector $s\indices{^{\xi}_{\alpha\xi}}$ states that the
covariant dynamics couples the scalar field $\phi$ to the torsion of spacetime.
Yet, the scalar field does not act as a source of torsion according to Eq.~(\ref{eq:feq-q2}).

The contravariant representation of the associated metric energy-momentum tensor~(\ref{eq:metric-em})
follows by virtue of
\begin{equation}\label{eq:g-deri1}
\pfrac{g}{g_{\mu\nu}}=g^{\nu\mu}\,g\qquad\Rightarrow\qquad
\pfrac{\sqrt{-g}}{g_{\mu\nu}}=\onehalf g^{\nu\mu}\sqrt{-g}
\end{equation}
as
\begin{align*}
T^{\mu\nu}&=\frac{2}{\sqrt{-g}}\pfrac{\HCdtz}{g_{\mu\nu}}\\
&=\pi^{*\,\mu}\pi^{\nu}+\pi^{*\,\nu}\pi^{\mu}-
g^{\mu\nu}\left(g_{\alpha\beta}\,\pi^{*\,\alpha}\pi^{\beta}-m^{2}\,\phi^{*}\phi\right).
\end{align*}
The canonical energy-momentum tensor for a system of complex scalar fields follows from the general prescription
\begin{align}
\theta\indices{^{\,\mu}_{\nu}}&=\pi^{*\,\mu}\pfrac{\HCd_{0}}{\pi^{*\,\nu}}+\pi^{\,\mu}\pfrac{\HCd_{0}}{\pi^{\,\nu}}
+\delta_{\nu}^{\mu}\left(\HCd_{0}-\pi^{*\,\alpha}\pfrac{\HCd_{0}}{\pi^{*\,\alpha}}-
\pi^{\,\alpha}\pfrac{\HCd_{0}}{\pi^{\,\alpha}}\right)\nonumber\\
&=\pi^{*\,\mu}\pi_{\nu}+\pi^{*}_{\nu}\,\pi^{\mu}-\delta_{\nu}^{\mu}\left(\pi^{*\,\alpha}\pi_{\alpha}-m^{2}\phi^{*}\phi\right).
\label{eq:csf-can-em1}
\end{align}
From Eq.~(\ref{eq:csf-can-em1}), the contravariant representation of the canonical energy-momentum tensor
for the complex Klein-Gordon system is obtained as
\begin{equation}\label{eq:KG-can-em}
\theta\indices{^{\,\mu\nu}}=\pi^{*\,\mu}\pi^{\nu}+\pi^{*\,\nu}\,\pi^{\,\mu}-
g^{\mu\nu}\left(g_{\alpha\beta}\,\pi^{*\,\alpha}\pi^{\,\beta}-m^{2}\phi^{*}\phi\right),
\end{equation}
which is symmetric and coincides with the above metric energy-momentum tensor of this system:
\begin{align*}
\theta^{\,\mu\nu}\equiv T^{\mu\nu}.
\end{align*}
Hence, for the system~(\ref{eq:kg-ham0}) there is no ambiguity with respect
to the source term of the generic equations~(\ref{eq:consistency1}) and~(\ref{eq:feq-k3}), which write for any $\HCdtD$
\begin{align*}
\pfrac{\HCdtD}{g_{\mu\nu}}&=-\pfrac{\HCdtz}{g_{\mu\nu}}\\
q\indices{_{\tau}^{\alpha\beta\lambda}}\pfrac{\HCdtD}{\tilde{q}\indices{_{\tau}^{\xi\beta\lambda}}}&=
\pfrac{\HCdtD}{\tilde{q}\indices{_{\alpha}^{\tau\beta\lambda}}}q\indices{_{\xi}^{\tau\beta\lambda}}\quad\Leftrightarrow\quad
R\indices{^{\eta}_{\tau\beta\lambda}}q\indices{^{\xi\tau\beta\lambda}}=
q\indices{^{\eta\tau\beta\lambda}}R\indices{^{\xi}_{\tau\beta\lambda}}.
\end{align*}
\subsection{Example~2: Proca system}
The Proca Hamiltonian in static spacetime was derived from the Proca Lagrangian
in Ref.\cite{StrRei12} by means of a (regular) Legendre transformation.
In a dynamic spacetime, the corresponding Proca Hamiltonian $\HCdtz\big(a_{\nu},\tilde{p}^{\nu\mu},g_{\mu\nu}\big)$ is given by
\begin{equation}\label{eq:Proca-Ham}
\HCdtz=-\quarter\tilde{p}^{\alpha\beta}\,\tilde{p}^{\xi\eta}g_{\alpha\xi}\,g_{\beta\eta}\frac{1}{\sqrt{-g}}-
\onehalf m^{2}a_{\alpha}\,a_{\beta}\,g^{\alpha\beta}\sqrt{-g},
\end{equation}
with the pertaining action integral
\begin{align}
S=\int_{R}\,\left[\,\onehalf\tilde{p}^{\,\alpha\beta}\left(a_{\alpha;\beta}-a_{\beta;\alpha}\right)+
\tilde{k}^{\,\alpha\lambda\beta}\,g_{\alpha\lambda;\beta}-\HCdtz\right]\dd^{4}x.
\label{eq:action-integral-Proca}
\end{align}
It follows from the variation of~(\ref{eq:action-integral-Proca}) that the
energy-momentum tensor is skew-symmetric, $\tilde{p}^{\alpha\beta}=-\tilde{p}^{\,\beta\alpha}$.
The covariant derivatives indicate that the system described by~(\ref{eq:action-integral-Proca})
is not closed, but depends on the connection coefficients $\gamma\indices{^{\,\nu}_{\beta\alpha}}$
as external functions of spacetime.
The diffeomorphism-invariant action integral of the closed system acquires
a modified form compared to that of Eq.~(\ref{eq:action-integral4}):
\begin{align}
S=\int_{R}\left[\onehalf\tilde{p}^{\,\alpha\beta}\left(a_{\alpha;\beta}-a_{\beta;\alpha}\right)+
\tilde{k}^{\,\alpha\lambda\beta}\,g_{\alpha\lambda;\beta}-
\onehalf\tilde{q}\indices{_{\eta}^{\alpha\xi\beta}}R\indices{^{\eta}_{\alpha\xi\beta}}-\HCdtD-\HCdtz\right]\dd^{4}x.
\label{eq:action-integral-Proca2}
\end{align}
By variation of~(\ref{eq:action-integral-Proca2}), the canonical equations for the Proca Ham\-iltonian~(\ref{eq:Proca-Ham})
follow from the general form of Eqs.~(\ref{eq:feq-vecfield}) as
\begin{align*}
\onehalf\left(a_{\mu;\nu}-a_{\nu;\mu}\right)&=\hphantom{-}\pfrac{\HCdtz}{\tilde{p}^{\mu\nu}}=-\onehalf p_{\mu\nu}\\
\tilde{p}\indices{^{\mu\beta}_{;\,\beta}}-2\tilde{p}^{\,\mu\beta}s\indices{^{\,\alpha}_{\beta\alpha}}
&=-\pfrac{\HCdtz}{a_{\mu}}=m^{2}a^{\mu}\sqrt{-g},
\end{align*}
hence, because of metric compatibility,
\begin{align}
p_{\mu\nu}&=\pfrac{a_{\nu}}{x^{\mu}}-\pfrac{a_{\mu}}{x^{\nu}}+2a_{\alpha}s\indices{^{\alpha}_{\mu\nu}}\nonumber\\
p\indices{^{\mu\beta}_{;\,\beta}}&=m^{2}a^{\mu}+2p^{\,\mu\beta}s\indices{^{\,\alpha}_{\beta\alpha}}.
\label{eq:proca-feq}
\end{align}
The vector field $a_{\mu}$ thus directly couples to the torsion of spacetime.
The other field equations, emerging from the variation of the action integral~(\ref{eq:action-integral-Proca2}),
are given by Eqs.~(\ref{eq:em-tensor-den}), (\ref{eq:feq-conn-coeff3}), and~(\ref{eq:feqs-Hdyn}).
Hence, the consistency equation retains the form of Eq.~(\ref{eq:consistency}).
From Eq.~(\ref{eq:u1-can-em0}), the canonical energy-momentum tensor is now obtained for this system as
\begin{equation}\label{eq:proca-can-em}
\theta\indices{^{\,\mu}_{\nu}}=-\onehalf p^{\,\alpha\mu}p_{\alpha\nu}+
\quarter\delta^{\mu}_{\nu}\left(p^{\alpha\beta}p_{\alpha\beta}-2m^{2}a^{\alpha}a_{\alpha}\right).
\end{equation}
Note that its covariant and contravariant representations are symmetric.
The metric energy-momentum tensor~(\ref{eq:metric-em}) can be set up considering Eq.~(\ref{eq:g-deri1}) and
\begin{align*}
\pfrac{g^{\alpha\beta}}{g_{\mu\nu}}=-g^{\alpha\mu}g^{\nu\beta}
\end{align*}
as
\begin{equation}\label{eq:proca-metric-em}
T\indices{^{\mu}_{\nu}}=-p^{\alpha\mu}p_{\alpha\nu}+m^{2}a^{\mu}a_{\nu}+
\quarter\delta^{\mu}_{\nu}\left(p^{\alpha\beta}p_{\alpha\beta}-2m^{2}a^{\alpha}a_{\alpha}\right).
\end{equation}
The difference of canonical and metric energy-momentum tensors is now
\begin{align*}
\theta\indices{^{\,\mu}_{\nu}}-T\indices{^{\mu}_{\nu}}&=\onehalf p^{\,\alpha\mu}p_{\alpha\nu}-m^{2}a^{\mu}a_{\nu}\\
&=-p^{\,\alpha\mu}\pfrac{\HCd_{0}}{p^{\alpha\nu}}+\pfrac{\HCd_{0}}{a_{\mu}}a_{\nu},
\end{align*}
hence
\begin{equation}\label{eq:proca-emt}
\theta\indices{^{\,\mu}_{\nu}}=T\indices{^{\mu}_{\nu}}-p^{\,\alpha\mu}\pfrac{\HCd_{0}}{p^{\alpha\nu}}+
\pfrac{\HCd_{0}}{a_{\mu}}a_{\nu}.
\end{equation}
The canonical energy-momentum tensor~(\ref{eq:proca-can-em}) thus enters into the
consistency equation~(\ref{eq:consistency1}) for the spacetime dynamics including torsion.

The contravariant representations of Eqs.~(\ref{eq:proca-can-em}) and~(\ref{eq:proca-metric-em}) for a Proca system are
\begin{alignat*}{2}
\theta^{\,\mu\nu}&=-\onehalf p^{\,\alpha\mu}p\indices{_{\alpha}^{\nu}}
&&+\quarter g^{\mu\nu}\left(p^{\alpha\beta}p_{\alpha\beta}-2m^{2}a^{\alpha}a_{\alpha}\right)\\
T^{\mu\nu}&=\,\,\,-p^{\,\alpha\mu}p\indices{_{\alpha}^{\nu}}+m^{2}a^{\mu}a^{\nu}&&+
\quarter g^{\mu\nu}\left(p^{\alpha\beta}p_{\alpha\beta}-2m^{2}a^{\alpha}a_{\alpha}\right).
\end{alignat*}
Our conclusion is that $\theta^{\,\mu\nu}$ represents the correct source term for a Proca system.
The \emph{canonical} energy-momentum tensor $\theta^{\,\mu\nu}$ thus entails an increased weighting of the kinetic energy
over the mass as compared to the \emph{metric} energy momentum tensor $T^{\mu\nu}$ in their roles as the source of gravity.
This holds independently of the particular model for the ``free'' (uncoupled)
gravitational field, whose dynamics is encoded in the Hamiltonian $\HCdtD$
in the generic Einstein-type equation~(\ref{eq:consistency1})
\begin{align*}
\theta^{\,\mu\nu}+\frac{2}{\sqrt{-g}}\pfrac{\HCdtD}{g_{\mu\nu}}=0.
\end{align*}
Due to the symmetry of the canonical energy-momentum tensor $\theta^{\,\mu\nu}$ of the Proca system,
the Hamiltonian $\HCdtD$ of the free gravitational field must be devised to entail
a correlation of the canonical momentum $q^{\,\mu\tau\beta\lambda}$ and the Riemann tensor
from the canonical equation~(\ref{eq:feqs-Hdyn}), which satisfies the condition from Eq.~(\ref{eq:feq-k3})
\begin{align*}
R\indices{^{\mu}_{\tau\beta\lambda}}q^{\nu\tau\beta\lambda}=q^{\,\mu\tau\beta\lambda}R\indices{^{\nu}_{\tau\beta\lambda}}.
\end{align*}
As will be shown in the following section, the metric energy-momentum tensor will turn out to be the appropriate source
term of gravity for systems invariant under the combination of the groups $\mathrm{Diff}(M)$ and $\mathrm{U}(1)$.
\section{System with U(1) symmetry\label{sec:u1dyn}}
\subsection{U(1) gauge theory\label{sec:u1}}
The action integral of a complex scalar field $\phi$ in conjunction with
a ``free'' real $4$-vector field $a_{\mu}$---which later acts as a ``gauge field''--- writes
\begin{align}
S=\int\left[\pi^{*\,\alpha}\pfrac{\phi}{x^{\alpha}}+\pfrac{\phi^{*}}{x^{\alpha}}\pi^{\alpha}+
p^{\,\alpha\beta}\pfrac{a_{\alpha}}{x^{\beta}}-\HCd_{0}\right]\dd^{4}x,
\label{eq:action-integral1}
\end{align}
with the initially uncoupled Hamiltonian $\HCd_{0}$ given by\cite{StrRei12,struckmeier17a}
\begin{equation}\label{eq:hd-kg-max}
\HCd_{0}=\pi^{*}_{\alpha}\pi^{\alpha}+m^{2}\phi^{*}\phi-\quarter p^{\alpha\beta}p_{\alpha\beta}.
\end{equation}
The action integral~(\ref{eq:action-integral1}) is obviously invariant
under the \emph{global} ($\Lambda=\mathrm{const.}$) symmetry transformation
\begin{align*}
p^{\nu\mu}&=P^{\nu\mu},&a_{\mu}&=A_{\mu}\\
\pi^{\mu}&=\Pi^{\mu}e^{-\rmi\Lambda},&\phi&=\Phi\,e^{-\rmi\Lambda}\\
\pi^{*\,\mu}&=\Pi^{*\,\mu}\,e^{\rmi\Lambda},&\phi^{*}&=\Phi^{*}\,e^{\rmi\Lambda}.
\end{align*}
The corresponding \emph{local}  ($\Lambda\neq\mathrm{const.}$) symmetry transformation
is defined by means of the generating function
\begin{equation}\label{eq:gen-erwet}
F_{2}^{\mu}=\Pi^{*\,\mu}\phi\,e^{\rmi\Lambda(x)}+
\phi^{*}\,\Pi^{\mu}e^{-\rmi\Lambda(x)}+P^{\alpha\mu}
\left(a_{\alpha}+\frac{1}{q}\pfrac{\Lambda}{x^{\alpha}}\right).
\end{equation}
In this context, the notation \emph{local} refers to the fact that
the generating function~(\ref{eq:gen-erwet}) depends \emph{explicitly} on $x^{\mu}$.
The general transformation rules applied to the actual generating function yield for the fields
\begin{align}
p^{\nu\mu}&=P^{\nu\mu},&a_{\mu}&=A_{\mu}-\frac{1}{q}\pfrac{\Lambda}{x^{\mu}}\nonumber\\
\pi^{\mu}&=\Pi^{\mu}e^{-\rmi\Lambda(x)},&\phi&=\Phi\,e^{-\rmi\Lambda(x)}\label{eq:rules}\\
\pi^{*\,\mu}&=\Pi^{*\,\mu}\,e^{\rmi\Lambda(x)},&\phi^{*}&=\Phi^{*}\,e^{\rmi\Lambda(x)}.\nonumber
\end{align}
The transformation rule for the Hamiltonians follows from the explicit
$x^{\mu}$-dependence of the generating function
\begin{align*}
\HCd^{\prime}=\HCd+\left.\pfrac{F_{2}^{\beta}}{x^{\beta}}\right|_{\mathrm{expl}},
\end{align*}
which means for the particular generating function~(\ref{eq:gen-erwet})
\begin{align*}
\left.\pfrac{F_{2}^{\beta}}{x^{\beta}}\right|_{\mathrm{expl}}=
\rmi\left(\Pi^{*\,\alpha}\phi\,e^{\rmi\Lambda(x)}-
\phi^{*}\,\Pi^{\alpha}e^{-\rmi\Lambda(x)}\right)\pfrac{\Lambda}{x^{\alpha}}+
\frac{1}{q}P^{\alpha\beta}\ppfrac{\Lambda}{x^{\alpha}}{x^{\beta}}.
\end{align*}
Inserting the inhomogeneous rule for the vector field $A_{\mu}$ yields
\begin{align}
\left.\pfrac{F_{2}^{\beta}}{x^{\beta}}\right|_{\mathrm{expl}}
&=\rmi q\left(\Pi^{*\,\alpha}\Phi-\Phi^{*}\,\Pi^{\alpha}\right)A_{\alpha}-
\rmi q\left(\pi^{*\,\alpha}\phi-\phi^{*}\,\pi^{\alpha}\right)a_{\alpha}\nonumber\\
&\quad+\onehalf P^{\alpha\beta}\left(\pfrac{A_{\alpha}}{x^{\beta}}+\pfrac{A_{\beta}}{x^{\alpha}}\right)-
\onehalf p^{\alpha\beta}\left(\pfrac{a_{\alpha}}{x^{\beta}}+\pfrac{a_{\beta}}{x^{\alpha}}\right).
\label{eq:KMG-ham-tra}
\end{align}
According to the canonical transformation formalism\cite{struckmeier17a},
it follows from Eq.~(\ref{eq:KMG-ham-tra}) that the uncoupled Hamiltonian~(\ref{eq:hd-kg-max})
is now replaced by the amended Hamiltonian $\HCd_{1}$, defined as
\begin{align}
\HCd_{1}=\pi_{\alpha}^{*}\pi^{\alpha}+\rmi q\left(\pi^{*\,\alpha}\phi-\phi^{*}
\pi^{\alpha}\right)a_{\alpha}+m^{2}\phi^{*}\phi-\quarter p^{\alpha\beta}p_{\alpha\beta}
+\frac{1}{2}p^{\alpha\beta}\left(\pfrac{a_{\alpha}}{x^{\beta}}+\pfrac{a_{\beta}}{x^{\alpha}}\right).
\label{eq:KMG-ham1}
\end{align}
This Hamiltonian $\HCd_{1}\big(\phi,\phi^{*},\pi^{\mu},\pi^{*\,\mu},a_{\nu},p^{\nu\mu}\big)$ is mapped under
the canonical transformation rules~(\ref{eq:rules}) of the fields and~(\ref{eq:KMG-ham-tra}) into a Hamiltonian
$\HCd_{1}\big(\Phi,\Phi^{*},\Pi^{\mu},\Pi^{*\,\mu},A_{\nu},P^{\nu\mu}\big)$
of exactly the same form in the transformed fields.

The Hamiltonian~(\ref{eq:KMG-ham1}) can finally be combined with the corresponding
terms in the initial action integral from Eq.~(\ref{eq:action-integral1}).
We thus end up with the particular action integral
\begin{align}
S=\int&\left[\pi^{*\,\alpha}\left(\pfrac{\phi}{x^{\alpha}}-\rmi q\,a_{\alpha}\right)+
\left(\pfrac{\phi^{*}}{x^{\alpha}}+\rmi q\,a_{\alpha}\right)\pi^{\alpha}
+\frac{1}{2}p^{\,\alpha\beta}\left(\pfrac{a_{\alpha}}{x^{\beta}}-
\pfrac{a_{\beta}}{x^{\alpha}}\right)-\HCd_{0}\right]\dd^{4}x.
\label{eq:action-integral1i}
\end{align}
The action integral~(\ref{eq:action-integral1i}) is \emph{form-invariant} under the local canonical transformation
rules~(\ref{eq:rules}) and~(\ref{eq:KMG-ham-tra}), which are generated by $F_{2}$ from Eq.~(\ref{eq:gen-erwet}).
It follows directly from the action integral~(\ref{eq:action-integral1i})
that $p^{\mu\nu}$ is skew-symmetric.
This property of the canonical momentum tensor conjugate to the vector field $a_{\mu}$
follows here from the gauge formalism and need not to be postulated.
Notice that the derivative of the vector field in Eq.~(\ref{eq:action-integral1})
is now replaced by a skew-symmetric tensor yielding a U$(1)$ gauge invariant action integral.
\subsection{Extension to the U(1)$\times$Diff(M) symmetry group}
The system of complex scalar and real vector fields described by the action~(\ref{eq:action-integral1i})
is form-invariant under a local $\mathrm{U}(1)$ symmetry transformation, hence under phase transformations
of the scalar field and the shift transformation of the real vector field.
In~\ref{app:u1xM}, we derive the combined transformation rules for the $\mathrm{U}(1)$ symmetry transformation
and the symmetry under the diffeomorphism group $\mathrm{Diff}(M)$ of the spacetime manifold $M$.
The Hamiltonian $\HCdtD\big(\tilde{k},\tilde{q},g\big)$ stands for the model which describes
both the dynamics of the momenta $\tilde{k}$ of the metric $g$, and $\tilde{q}$
the momenta of the connection coefficients $\gamma$.
On the basis of the action~(\ref{eq:action-integral1i}), the generally covariant action
which describes in addition the interaction of the massive complex scalar field $\phi$
and the massless real (Maxwell) vector field $a_{\mu}$ with the spacetime geometry is obtained as
\begin{align}
S=\int_{R}\,&\Bigg[\tilde{\pi}^{*\,\alpha}\left(\pfrac{\phi}{x^{\alpha}}-\rmi q\,a_{\alpha}\phi\right)+
\left(\pfrac{\phi^{*}}{x^{\alpha}}+\rmi q\,\phi^{*}a_{\alpha}\right)\tilde{\pi}^{\alpha}\nonumber\\
&\mbox{}+\frac{1}{2}\tilde{p}^{\,\alpha\beta}\left(\pfrac{a_{\alpha}}{x^{\beta}}-\pfrac{a_{\beta}}{x^{\alpha}}\right)
+\tilde{k}^{\,\alpha\lambda\beta}\,g_{\alpha\lambda;\,\beta}-
\frac{1}{2}\tilde{q}\indices{_{\xi}^{\lambda\alpha\beta}}R\indices{^{\,\xi}_{\lambda\alpha\beta}}-\HCdtz-\HCdtD\Bigg]\dd^{4}x.
\label{eq:S_final}
\end{align}
Clearly, this system is form-invariant under the combined symmetry group $\mathrm{U}(1)\times\mathrm{Diff}(M)$.
Comparing the invariant action functional~(\ref{eq:S_final}) with that of the Proca system (Eq.~(\ref{eq:action-integral-Proca2}))
shows that now the gauge field $a_{\mu}$ does no longer couple directly to the connection $\gamma\indices{^{\xi}_{\alpha\beta}}$.
The Proca action contains the additional term
\begin{align*}
\onehalf\tilde{p}^{\alpha\beta}\left(\gamma\indices{^{\xi}_{\alpha\beta}}-\gamma\indices{^{\xi}_{\beta\alpha}}\right)a_{\xi},
\end{align*}
which is invariant under the $\mathrm{Diff}(M)$ symmetry group, but not under the group $\mathrm{U}(1)\times\mathrm{Diff}(M)$.

With $\HCdtz$ in particular the sum of the Klein-Gordon Hamiltonian~(\ref{eq:kg-ham0}) and the
massless Proca Hamiltonian~(\ref{eq:Proca-Ham}), Eq.~(\ref{eq:S_final}) represents the generic
action of the Einstein's ``unified field theory'' of electromagnetics and gravitation.
The Hamiltonian $\HCdtD\big(\tilde{k},\tilde{q},g\big)$ stands for all
possible descriptions of the free gravitational field---which are not restricted to metric compatibility and zero torsion.
The final ``gauged'' Hamiltonian is thus with $\HCd_{0}$ from Eq.~(\ref{eq:hd-kg-max})
\begin{align}
\tilde{\HCd}_{3}&=\HCdtz+\HCdtD+
\rmi q\left(\tilde{\pi}^{*\,\alpha}\phi-\phi^{*}\tilde{\pi}^{\alpha}\right)a_{\alpha}\nonumber\\
&\quad+\frac{1}{2}\tilde{p}^{\alpha\beta}\left(\pfrac{a_{\alpha}}{x^{\beta}}+
\pfrac{a_{\beta}}{x^{\alpha}}\right)+\left(\tilde{k}^{\alpha\lambda\beta}g_{\xi\lambda}+
\tilde{k}^{\lambda\alpha\beta}g_{\lambda\xi}\right)\gamma\indices{^{\xi}_{\alpha\beta}}\nonumber\\
&\quad+\frac{1}{2}\tilde{q}\indices{_{\eta}^{\alpha\xi\beta}}\left(
\pfrac{\gamma\indices{^{\eta}_{\alpha\xi}}}{x^{\beta}}+\pfrac{\gamma\indices{^{\eta}_{\alpha\beta}}}{x^{\xi}}+
\gamma\indices{^{\kappa}_{\alpha\beta}}\gamma\indices{^{\eta}_{\kappa\xi}}-
\gamma\indices{^{\kappa}_{\alpha\xi}}\gamma\indices{^{\eta}_{\kappa\beta}}\right),
\label{eq:Hamilton2}
\end{align}
from which the canonical field equations follow now as tensor equations.
\subsection{Canonical Field equations and the consistency equation}
In this section, we derive the set of canonical field equations emerging
from the gauge-invariant Hamiltonian $\tilde{\HCd}_{3}$ of Eq.~(\ref{eq:Hamilton2}).
To begin with, the field equations for the complex scalar field $\phi$ and its conjugates follow as
\begin{align*}
\pfrac{\phi}{x^{\mu}}&=\hphantom{-}\pfrac{\tilde{\HCd}_{3}}{\tilde{\pi}^{*\,\mu}}=
\hphantom{-}\pfrac{\tilde{\HCd}_{0}}{\tilde{\pi}^{*\,\mu}}-\rmi q\,a_{\mu}\phi\\
\pfrac{\phi^{*}}{x^{\mu}}&=\hphantom{-}\pfrac{\tilde{\HCd}_{3}}{\tilde{\pi}^{\mu}}=
\hphantom{-}\pfrac{\tilde{\HCd}_{0}}{\tilde{\pi}^{\mu}}+\rmi q\,\phi^{*}a_{\mu}\\
\pfrac{\tilde{\pi}^{\alpha}}{x^{\alpha}}&=-\pfrac{\tilde{\HCd}_{3}}{\phi^{*}}=
-\pfrac{\tilde{\HCd}_{0}}{\phi^{*}}-\rmi q\,\tilde{\pi}^{*\,\alpha}a_{\alpha}\\
\pfrac{\tilde{\pi}^{*\,\alpha}}{x^{\alpha}}&=-\pfrac{\tilde{\HCd}_{3}}{\phi}=
-\pfrac{\tilde{\HCd}_{0}}{\phi}+\rmi q\,\tilde{\pi}^{\alpha}a_{\alpha}.
\end{align*}
The field equations related to the real vector field $a_{\mu}$ emerge as
\begin{align*}
\pfrac{a_{\nu}}{x^{\mu}}&=\hphantom{-}\pfrac{\tilde{\HCd}_{3}}{\tilde{p}^{\nu\mu}}=
\pfrac{\tilde{\HCd}_{0}}{\tilde{p}^{\nu\mu}}+\frac{1}{2}\left(\pfrac{a_{\nu}}{x^{\mu}}+\pfrac{a_{\mu}}{x^{\nu}}\right)\\
\pfrac{\tilde{p}^{\,\nu\beta}}{x^{\beta}}&=-\pfrac{\tilde{\HCd}_{3}}{a_{\nu}}=
-\pfrac{\tilde{\HCd}_{0}}{a_{\nu}}-\rmi q\left(\tilde{\pi}^{*\,\nu}\phi-\phi^{*}\tilde{\pi}^{\nu}\right),
\end{align*}
hence
\begin{align*}
\pfrac{\tilde{\HCd}_{0}}{\tilde{p}^{\nu\mu}}&=\frac{1}{2}\left(\pfrac{a_{\nu}}{x^{\mu}}-\pfrac{a_{\mu}}{x^{\nu}}\right)\\
\pfrac{\tilde{\HCd}_{0}}{a_{\nu}}&=-\pfrac{\tilde{p}^{\,\nu\beta}}{x^{\beta}}
-\rmi q\left(\tilde{\pi}^{*\,\nu}\phi-\phi^{*}\tilde{\pi}^{\nu}\right).
\end{align*}
For metric compatibility, hence for $\HCdtD$ not depending on $\tilde{k}\indices{^{\alpha\lambda\beta}}$,
the field equations related to the (symmetric) metric tensor $g_{\mu\nu}$ are
\begin{align*}
\pfrac{g_{\alpha\lambda}}{x^{\beta}}&=\hphantom{-}\pfrac{\tilde{\HCd}_{3}}{\tilde{k}\indices{^{\alpha\lambda\beta}}}
=g_{\kappa\lambda}\gamma\indices{^{\kappa}_{\alpha\beta}}+g_{\alpha\kappa}\gamma\indices{^{\kappa}_{\lambda\beta}}
\quad\Leftrightarrow\quad g_{\alpha\lambda\,;\,\beta}\equiv0\\
\pfrac{\tilde{k}^{\xi\lambda\beta}}{x^{\beta}}&=-\pfrac{\tilde{\HCd}_{3}}{g_{\xi\lambda}}
=-\tilde{k}^{\alpha\lambda\beta}\gamma\indices{^{\xi}_{\alpha\beta}}-
\tilde{k}^{\xi\alpha\beta}\gamma\indices{^{\lambda}_{\alpha\beta}}-\pfrac{\tilde{\HCd}_{0}}{g_{\xi\lambda}}-
\pfrac{\HCdtD}{g_{\xi\lambda}}.
\end{align*}
Finally, the field equations for the connection $\gamma\indices{^{\eta}_{\alpha\xi}}$ and its conjugate,
$\tilde{q}\indices{_{\eta}^{\alpha\xi\beta}}$, are set up:
\begin{align*}
\pfrac{\gamma\indices{^{\eta}_{\alpha\xi}}}{x^{\beta}}&=
\pfrac{\tilde{\HCd}_{3}}{\tilde{q}\indices{_{\eta}^{\alpha\xi\beta}}}\\
&=\pfrac{\HCdtD}{\tilde{q}\indices{_{\eta}^{\alpha\xi\beta}}}+\frac{1}{2}\left(
\pfrac{\gamma\indices{^{\eta}_{\alpha\xi}}}{x^{\beta}}+\pfrac{\gamma\indices{^{\eta}_{\alpha\beta}}}{x^{\xi}}+
\gamma\indices{^{\kappa}_{\alpha\beta}}\gamma\indices{^{\eta}_{\kappa\xi}}-
\gamma\indices{^{\kappa}_{\alpha\xi}}\gamma\indices{^{\eta}_{\kappa\beta}}\right),
\end{align*}
which yields the negative Riemann tensor $R\indices{^{\eta}_{\alpha\xi\beta}}$
\begin{align}
2\pfrac{\HCdtD}{\tilde{q}\indices{_{\eta}^{\alpha\xi\beta}}}&=
\pfrac{\gamma\indices{^{\eta}_{\alpha\xi}}}{x^{\beta}}-
\pfrac{\gamma\indices{^{\eta}_{\alpha\beta}}}{x^{\xi}}+
\gamma\indices{^{\kappa}_{\alpha\xi}}\gamma\indices{^{\eta}_{\kappa\beta}}-
\gamma\indices{^{\kappa}_{\alpha\beta}}\gamma\indices{^{\eta}_{\kappa\xi}}\nonumber\\
&=-R\indices{^{\eta}_{\alpha\xi\beta}}.
\label{eq:gamma-deri}
\end{align}
As $\HCdtz$ and $\HCdtD$ do not depend on the connection $\gamma\indices{^{\xi}_{\alpha\beta}}$
the equation for its conjugate, $\tilde{q}\indices{_{\xi}^{\alpha\beta\mu}}$, is fully determined
\begin{equation}\label{eq:feq-div-q}
\pfrac{\tilde{q}\indices{_{\xi}^{\alpha\beta\lambda}}}{x^{\lambda}}=
-\pfrac{\tilde{\HCd}_{3}}{\gamma\indices{^{\xi}_{\alpha\beta}}}=2\tilde{k}^{\,\lambda\alpha\beta}g_{\lambda\xi}+
\tilde{q}\indices{_{\eta}^{\alpha\beta\lambda}}\gamma\indices{^{\eta}_{\xi\lambda}}+
\tilde{q}\indices{_{\xi}^{\eta\lambda\beta}}\gamma\indices{^{\alpha}_{\eta\lambda}}.
\end{equation}
Similarly to the corresponding field equation of the pure diffemorphism symmetry
from Eq.~(\ref{eq:feq-conn-coeff3}), we can rewrite this equation in terms of a covariant divergence
\begin{equation}\label{eq:feq-conn-coeff5}
\tilde{q}\indices{_{\eta}^{\xi\lambda\beta}_{;\,\beta}}=
-2g_{\eta\beta}\tilde{k}^{\,\beta\xi\lambda}+
\tilde{q}\indices{_{\eta}^{\xi\beta\alpha}}s\indices{^{\,\lambda}_{\beta\alpha}}+
2\tilde{q}\indices{_{\eta}^{\xi\lambda\beta}}s\indices{^{\alpha}_{\beta\alpha}}.
\end{equation}
The term $a_{\eta}\tilde{p}^{\xi\lambda}$ is now missing as the vector field does not
couple to the connection in the Hamiltonian $\tilde{\HCd}_{3}$.
Consequently, the consistency equations for metric compatibility now follow as
\begin{equation}
-\pfrac{\HCdtD}{g_{\mu\nu}}=\pfrac{\HCdtz}{g_{\mu\nu}},
\label{eq:consistency-u1}
\end{equation}
and
\begin{align*}
q\indices{_{\tau}^{\alpha\beta\lambda}}\pfrac{\HCdtD}{\tilde{q}\indices{_{\tau}^{\xi\beta\lambda}}}&=
\pfrac{\HCdtD}{\tilde{q}\indices{_{\alpha}^{\tau\beta\lambda}}}q\indices{_{\xi}^{\tau\beta\lambda}}\\
q\indices{^{\eta\xi\lambda\beta}_{;\,\beta}}&=
q\indices{^{\eta\xi\beta\alpha}}s\indices{^{\,\lambda}_{\beta\alpha}}+
2q\indices{^{\eta\xi\lambda\beta}}s\indices{^{\alpha}_{\beta\alpha}}.
\end{align*}
In contrast to the $\mathrm{Diff}(M)$ gauge theory of the Proca system, for which the \emph{canonical}
energy-momentum tensor emerges as the source in the consistency equation~(\ref{eq:consistency}),
the $\mathrm{U}(1)\times\mathrm{Diff}(M)$ gauge theory has the \emph{metric} energy-momentum tensor as its source.
For the conventional case of $\tilde{\HCd}_{0}$ not depending on the conjugate of the metric,
$\tilde{k}^{\alpha\lambda\mu}$, this tensor is generally obtained according to Eq.~(\ref{eq:metric-em}).
For the Klein-Gordon-Maxwell Hamiltonian~(\ref{eq:KMG-ham1}), hence for the $\mathrm{U}(1)$ gauge theory,
this tensor has the explicit Hamiltonian form
\begin{align}
T^{\mu\nu}=\pi^{*\,\mu}\pi^{\nu}+\pi^{*\,\nu}\,\pi^{\mu}-p^{\,\mu\alpha}p^{\nu\beta}g_{\alpha\beta}
-g^{\mu\nu}\left(\pi^{*}_{\alpha}\,\pi^{\alpha}-m^{2}\,\phi^{*}\,\phi-\quarter p^{\alpha\beta}p_{\alpha\beta}\right).
\label{eq:metric-em3}
\end{align}
It can be shown similarly that the \emph{metric} energy momentum tensor also acts as the
source term for $\mathrm{SU}(N)\times\mathrm{Diff}(M)$ gauge theories, hence for the
general relativistic extension of Yang-Mills gauge theories.
This again holds independently of the particular $\HCdtD$ describing the dynamics of the ``free'' gravitational field.
\section{Conclusions\label{sec:conclusions}}
Any (globally) Lorentz-invariant Lagrangian/Hamiltonian system can be converted into an
amended Lagrangian or Hamiltonian, which is form-invariant under the $\mathrm{Diff}(M)$
symmetry group, following the well-established reasoning of gauge theories.
The integrand of the final action integral~(\ref{eq:action-integral4}) was shown to represent a proper (world) scalar density,
thereby meeting the requirement of Einstein's \emph{Principle of General Relativity}, i.e.\
form-invariance under \emph{local} chart transitions (diffeomorphisms).
For simplicity, metric compatibility was imposed later in Eq.~(\ref{eq:metricity}).

The gauge formalism reveals that scalar and vector source fields couple differently to a dynamic spacetime.
In the context of this paper, we must distinguish three cases:
\begin{enumerate}
\item Massive or massless, real or complex scalar fields $\phi$ are associated with the metric (Hilbert)
energy-momentum tensor $T^{\mu\nu}$ as the source term of the Einstein-type equation~(\ref{eq:consistency1}).
Yet, in this case, $T^{\mu\nu}$  coincides with the canonical energy-momentum tensor $\theta^{\mu\nu}$.
Scalar fields do not act as source for a torsion of spacetime.
\item Systems of massive or non-massive vector fields $a_{\mu}$ with no other symmetries
than the $\mathrm{Diff}(M)$ group---such as the Proca system---require the canonical
energy-momentum tensor~$\theta^{\mu\nu}$ as the source term.
Such systems do couple to a torsion of spacetime.
\item A system consisting of a complex (charged) scalar field $\phi$, which couples minimally to a
massless vector field $a_{\mu}$ (Maxwell field)---hence a system with additional $\mathrm{U}(1)$ sym\-metry---has
the metric energy-momentum tensor $T^{\mu\nu}$ as the source term.
This can be generalized to systems with $\mathrm{SU}(N)$ symmetry\cite{strgrei19}.
These systems do not couple to torsion of spacetime.
\end{enumerate}
The general prescription to promote a Lorentz-invariant action into a generally covariant action is thus
to replace the partial derivatives of all non-scalar objects in the action integral by covariant derivatives.
Exceptions to this recipe are twofold.
The first exception is encountered regarding the invariant action integral~(\ref{eq:action-integral1i})
of the $\mathrm{U}(1)$ gauge theory.
The direct coupling $\tilde{p}^{\alpha\beta}a_{\xi}\gamma\indices{^{\xi}_{\alpha\beta}}$ to the dynamical spacetime geometry
is exactly compensated due to the inhomogeneous transformation rule~(\ref{eq:trafoa}) in Eq.~(\ref{eq:gf-p2}).
The remaining derivative of $\HCdtz$ with respect to the metric $g_{\alpha\beta}$
then yields the metric energy-momentum tensor as the appropriate source term for the
spacetime dynamics of a system with $\mathrm{U}(1)$ resp.\ $\mathrm{SU}(N)$ symmetry.
Moreover, due to the missing coupling term, the canonical spin tensor $\tau^{\eta\xi\lambda}$
does not show up on the right-hand side of Eq.~(\ref{eq:feq-q2}).
Hence, the system does not generate torsion of spacetime.

The second exception is the partial derivative of the affine connection in the action integral.
As the affine connection is no tensor, its partial derivatives cannot directly be
converted into covariant derivatives in the initial action integral~(\ref{eq:action-integral2}).
Yet, by virtue of the gauge formalism, a term quadratic in the affine connection $\gamma$ emerges,
which is shown to make the partial deri\-vatives of $\gamma\indices{^{\eta}_{\alpha\beta}}$
into one-half the Riemann tensor---and hence into a generally covariant object.
The action integral~(\ref{eq:action-integral4}) thus complies with the \emph{Principle of General Relativity}.
Its subsequent field equations are then tensor equations, which quantify the interaction of the source fields
with the spacetime geometry, the latter being described by the metric and the affine connection
as separate geometrical objects.
The canonical transformation approach to spacetime dynamics thus naturally implements the Palatini formalism\cite{palatini19}.

As the gauge formalism determines only the coupling of matter and spacetime dynamics,
an additive Hamiltonian $\HCdtD$ of the ``free'' gravitational field is to be postulated.
An action with a quadratic term in the canonical momenta of the gauge field is required
to obtain a closed system of field equations for the coupled dynamics of fields and spacetime geometry\cite{struckmeier17a}.

We have shown in this paper that the correct energy-mo\-mentum tensor for the massive vector is the canonical one,
whereas the metric one is correct for massless vector fields, hence for systems with additional U(1) or, more general, SU(N) symmetry.
As a consequence, compact astrophysical objects, like neutron stars and binary neutron star mergers
must be reinvestigated with the appropriate canonical energy-momentum terms for the vector repulsion from an effective field theory (EFT).
A similar conclusion does also hold for fermions, both for protons and electrons, as well as for neutrinos,
both in white dwarfs, neutron stars and in ``ultra high energy cosmic ray'' (UHECR) events.
The particular coupling of \emph{spinor fields} with spacetime will be the topic of separate papers\cite{vasak19,struckmeier19}.
It will be shown that the coupling of a spinor with the spacetime dynamics
gives rise to an effective mass term in the generally covariant Dirac equation.

\smallskip
The theory of geometrodynamics with quadratic Riemann tensor and a modified coupling of vector fields developed here entails important
astrophysical and cosmological consequences according to the new general equation~(61) of Ref.\cite{struckmeier17a}.
Details will be addressed in a forthcoming article, yet qualitative consequences are discussed briefly in the following.
Reference\cite{struckmeier17a} demonstrated that Eq.~(61) is solved as
well as the classical Einstein equation by the Schwarzschild and the Kerr metrics.
However, due to the canonical quadratic curvature terms, a new understanding emerges of both Friedman cosmology,
as well as for the interior of compact stellar objects, pulsar dynamics and binary neutron star mergers.
The canonical energy-momentum tensor replaces the Einstein's metric tensor, resulting in
structural changes of compact astrophysical objects and relativistic collapse dynamics.

The interior structure of the neutron stars can be calculated using the QCD-motivated relativistic
mean-field (RMF) theory with meson fields interacting with neutrons, other baryons, and with leptons.
This is done e.g.\ in the Walecka \cite{walecka74,serot86} and in non-linear Boguta-Bodmer
RMF theories \cite{boguta77,steinheimer11} by introducing the repulsive $\omega^{\mu}$ and $\rho^{\mu}$ interactions,
as well as the attractive scalar $\sigma$ field.
For both $N=Z$ nuclear matter, and for the $N$-dominated neutron matter, the nucleons and heavier baryons
acquire substantial relativistic effects in the dense medium, including both, strong attraction
($U_{s}\simeq -450$~MeV) induced by scalar $\sigma$~mesons with mass $m_{\sigma}\simeq 550$~MeV,
and strong repulsion ($U_v\simeq 350$~MeV) due to the $\omega^{\mu}$ and $\rho^{\mu}$~vector mesons
with masses $m_{\omega}\simeq m_{\rho}$ of the order of $800$~MeV.
Due to scalar-meson interactions, the effective in-medium nucleon mass drops from the value
$m_{N}\simeq 940$~MeV to about $m_{N}^{*}\simeq 550$~MeV at densities around the nuclear ground state density $\rho_{0}$.
In this case contributions of massive meson vector (spin-$1$) fields to the energy density
are of the same order of magnitude as the energy density of neutrons and other baryons,
exceeding the baryon kinetic energy density above baryon densities of $2\rho_{0}\simeq 0.3$~fm$^{-3}$.

Neutron star formations in supernovae collapses, and creation of hypermassive neutron stars (HMNS)
in binary neutron star mergers, may produce even larger relativistic effects
due to strong magnetic fields in pulsars, in rapidly rotating neutron stars, and even more in violently
moving binary HMNSs with high temperatures, higher densities
and strong magnetic fields $H\simeq 10^{18}$~Gauss.
The high angular frequencies may align spins of both, nucleons (spin-$1/2$) and vector mesons, with similar masses.
Hence, the terms proportional to $p^{\mu\beta}p\indices{^{\nu}_{\beta}}$ and $a^{\mu}a^{\nu}$
on the right-hand side of Eq.~(\ref{eq:proca-emt}) can be of the same order of magnitude
as the metric energy-momentum tensor $T^{\mu\nu}$ of the Einstein equation.
On this account, the theory of neutron star structure and dynamics need to be fundamentally revised.

The recent detection of the gravitational wave from a binary neutron star merger
by the LIGO-VIRGO collaboration (GW170817)\cite{coulter17,abbott2017gw170817} opens various
interesting astrophysical scenarios: analysis of the gravitational wave data,
in combination with the independently detected gamma-ray burst (GRB 170817A)\cite{abbott2017gravitational}
and further electromagnetic radiation\cite{abbott2017multi} results in a neutron
star merger scenario which is in good agreement with numerical simulations of binary
neutron star mergers performed in full general relativistic hydrodynamics.
As a result of the binary merger, a fast, differentially rotating HMNS was produced,
which lived for $\approx 1$ second before it collapsed to a rotating Kerr black hole.
Matter in the interior of the HMNS reaches densities of
up to several times normal nuclear matter, and temperatures could reach $T\sim 50-100$~MeV.
However, for such high densities, the equation of state (EOS) is still poorly constrained.
High energy heavy ion collision data are compatible with a hadron-quark phase transition,
which then shall also be present in the interior of the HMNS \cite{hanauske2017concluding}.

As predicted by Csernai and coworkers\cite{csernai13}, a strongly rotating quark-gluon plasma
has been detected experimentally in non-central ultra-relativistic heavy ion collisions\cite{adamczyk2017global}.
The coupling of the internal spin component of the quark-gluon plasma phase
(in the interior of the HMNS) may act as a source of torsion of spacetime.
Therefore, numerical simulations of the post-merger phase of a neutron star merger performed
within the canonical theory of gravitation will differ fundamentally from classical Einstein simulations.

The replacement of the metric energy-momentum-tensor by the canonical energy-momentum-tensor
thus results in two important changes in the general relativistic descriptions of neutron stars
and binary neutron star mergers: Both the static Tolman-Oppenheimer-Volkov equation and the dynamic theory of
binary neutron stars become invalid, because not only the EOS of hot, dense, and strongly rotating matter,
but also the metric itself will look quite differently as compared to the Einstein general relativity.
\section*{Acknowledgements}
The authors are deeply indebted to the ``Walter Greiner-Gesell\-schaft zur F\"{o}rderung
der physikalischen Grundlagenforschung e.V.'' in Frankfurt for its support.
H.~Stoecker acknowledges the Judah M.~Eisenberg Professor Laureatus Chair at the Institute
or Theoretical Physics of the Goethe University Frankfurt.
\appendix
\section{Proof of the consistency equation~(\ref{eq:consistency})\label{app:ricci-formula}}
If a dynamical quantity $t\indices{^{\alpha_{1}\ldots\alpha_{n}}_{\beta_{1}\ldots\beta_{m}}}(x)$ at the spacetime
location~$x$ transforms to $T\indices{^{\xi_{1}\ldots\xi_{n}}_{\eta_{1}\ldots\eta_{m}}}(X)$ at $X$ according to
\begin{equation}\label{eq:def-tensor-denity}
T\indices{^{\xi_{1}\ldots\xi_{n}}_{\eta_{1}\ldots\eta_{m}}}=
t\indices{^{\alpha_{1}\ldots\alpha_{n}}_{\beta_{1}\ldots\beta_{m}}}
\pfrac{X^{\xi_{1}}}{x^{\alpha_{1}}}\ldots\pfrac{X^{\xi_{n}}}{x^{\alpha_{n}}}
\pfrac{x^{\beta_{1}}}{X^{\eta_{1}}}\ldots\pfrac{x^{\beta_{m}}}{X^{\eta_{m}}}
\left|\pfrac{x}{X}\right|^{w}
\end{equation}
with $\left|\pfrac{x}{X}\right|$ the determinant of the Jacobi matrix (``Jacobian'') of the transformation $x\mapsto X$
\begin{align*}
\left|\pfrac{x}{X}\right|=\pfrac{\left(x^{0},\ldots,x^{3}\right)}{\left(X^{0},\ldots,X^{3}\right)},
\end{align*}
then $T$ is referred to as a \emph{relative $(n,m)$-tensor of weight} $w$.
The difference of the second covariant derivatives of this kind of tensor
is given by the Ricci formula\cite{plebanski06}\footnote{Note that Ref.\cite{plebanski06}
in its Eq.~(3.10) uses the opposite sign convention in the definition
of the weight factor $w$, as compared to Eq.~(\ref{eq:def-tensor-denity}).
In our case, $\sqrt{-g}$ represents a relative scalar of weight $w=+1$.
So all momentum fields in our paper follow as relative tensors of weight $w=+1$.}
\begin{align}
&\quad\left(T\indices{^{\xi_{1}\ldots\xi_{n}}_{\eta_{1}\ldots\eta_{m}}}\right)_{;\,\mu\,;\,\nu}-
\left(T\indices{^{\xi_{1}\ldots\xi_{n}}_{\eta_{1}\ldots\eta_{m}}}\right)_{;\,\nu\,;\,\mu}\nonumber\\
&=\sum_{k=1}^{m}R\indices{^{\tau}_{\eta_{k}\mu\nu}}T\indices{^{\xi_{1}\ldots\xi_{n}}_{\eta_{1}\ldots\tau\ldots\eta_{m}}}-
\sum_{k=1}^{n}R\indices{^{\,\xi_{k}}_{\tau\mu\nu}}T\indices{^{\xi_{1}\ldots\tau\ldots\xi_{n}}_{\eta_{1}\ldots\eta_{m}}}\nonumber\\
&\quad+wR\indices{^{\tau}_{\tau\mu\nu}}T\indices{^{\xi_{1}\ldots\xi_{n}}_{\eta_{1}\ldots\eta_{m}}}-
2s\indices{^{\tau}_{\mu\nu}}\left(T\indices{^{\xi_{1}\ldots\xi_{n}}_{\eta_{1}\ldots\eta_{m}}}\right)_{\,;\,\tau},
\label{eq:ricci-formula}
\end{align}
where $R\indices{^{\,\xi}_{\tau\mu\nu}}$ denotes the Riemann curvature tensor~(\ref{eq:riemann-tensor})
and $s\indices{^{\tau}_{\mu\nu}}$ the torsion tensor~(\ref{eq:torsion-tensor}).
In the first and second term on the right-hand side of Eq.~(\ref{eq:ricci-formula}),
the index $\tau$ is located at the $k$-th position of the index list, respectively.

From the definition of the Riemann tensor~(\ref{eq:riemann-tensor}), its contraction
$R\indices{^{\tau}_{\tau\beta\lambda}}$ immediately follows as
\begin{equation}\label{eq:Rie-contraction}
R\indices{^{\tau}_{\tau\beta\lambda}}=\pfrac{\gamma\indices{^{\tau}_{\tau\lambda}}}{x^{\beta}}-
\pfrac{\gamma\indices{^{\tau}_{\tau\beta}}}{x^{\lambda}}.
\end{equation}
If the Riemann tensor is assumed to be skew-symmetric in its \emph{first} index pair, then
\begin{align*}
R\indices{^{\tau}_{\tau\beta\lambda}}=R\indices{_{\eta\tau\beta\lambda}}g^{\eta\tau}=0.
\end{align*}
Yet, we do not pursue the assumption at this point in order to maintain the consistency of our
derivation, where we did not discuss symmetry properties of Eq.~(\ref{eq:feq-conn-coeff3}).

With the Ricci formula in the general form of Eq.~(\ref{eq:ricci-formula}), we calculate
the second covariant derivative $\tilde{q}\indices{_{\eta}^{\xi\lambda\beta}_{;\beta;\lambda}}$
of the $(3,1)$-tensor $\tilde{q}$ of weight $w=1$ from Eq.~(\ref{eq:feq-conn-coeff3}).
Owing to the skew-symmetry of $\tilde{q}\indices{_{\eta}^{\xi\lambda\beta}}$ in its last index pair,
$\lambda$ and $\beta$, the second covariant derivative can be expressed as a difference
of two second covariant derivatives with the differentiation sequence reversed.
Then the Ricci formula~(\ref{eq:ricci-formula}) can be applied
\begin{align}
2\tilde{q}\indices{_{\eta}^{\xi\lambda\beta}_{;\beta;\lambda}}&=
\tilde{q}\indices{_{\eta}^{\xi\lambda\beta}_{;\beta;\lambda}}-
\tilde{q}\indices{_{\eta}^{\xi\beta\lambda}_{;\beta;\lambda}}\nonumber\\
&=\tilde{q}\indices{_{\eta}^{\xi\lambda\beta}_{;\beta;\lambda}}-
\tilde{q}\indices{_{\eta}^{\xi\lambda\beta}_{;\lambda;\beta}}\nonumber\\
&=R\indices{^{\tau}_{\eta\beta\lambda}}\tilde{q}\indices{_{\tau}^{\xi\lambda\beta}}-
\tilde{q}\indices{_{\eta}^{\tau\lambda\beta}}R\indices{^{\xi}_{\tau\beta\lambda}}-
R\indices{^{\lambda}_{\tau\beta\lambda}}\tilde{q}\indices{_{\eta}^{\xi\tau\beta}}-
R\indices{^{\beta}_{\tau\beta\lambda}}\tilde{q}\indices{_{\eta}^{\xi\lambda\tau}}\nonumber\\
&\quad\mbox{}+R\indices{^{\tau}_{\tau\beta\lambda}}\tilde{q}\indices{_{\eta}^{\xi\lambda\beta}}-
2s\indices{^{\tau}_{\beta\lambda}}\tilde{q}\indices{_{\eta}^{\xi\lambda\beta}_{;\tau}}\nonumber\\
&=R\indices{^{\tau}_{\eta\beta\lambda}}\tilde{q}\indices{_{\tau}^{\xi\lambda\beta}}-
\tilde{q}\indices{_{\eta}^{\tau\lambda\beta}}R\indices{^{\xi}_{\tau\beta\lambda}}-
2s\indices{^{\tau}_{\beta\lambda}}\tilde{q}\indices{_{\eta}^{\xi\lambda\beta}_{;\tau}}\nonumber\\
&\quad\mbox{}+\left(R\indices{^{\beta}_{\tau\beta\lambda}}+
R\indices{^{\beta}_{\beta\lambda\tau}}+R\indices{^{\beta}_{\lambda\tau\beta}}\right)
\tilde{q}\indices{_{\eta}^{\xi\tau\lambda}}\nonumber\\
&=R\indices{^{\tau}_{\eta\beta\lambda}}\tilde{q}\indices{_{\tau}^{\xi\lambda\beta}}-
\tilde{q}\indices{_{\eta}^{\tau\lambda\beta}}R\indices{^{\xi}_{\tau\beta\lambda}}-
2s\indices{^{\tau}_{\beta\lambda}}\tilde{q}\indices{_{\eta}^{\xi\lambda\beta}_{;\tau}}\nonumber\\
&\quad\mbox{}+4\left(s\indices{^{\beta}_{\tau\beta;\lambda}}+\onehalf
s\indices{^{\beta}_{\lambda\tau;\beta}}+s\indices{^{\beta}_{\sigma\beta}}
s\indices{^{\sigma}_{\tau\lambda}}\right)\tilde{q}\indices{_{\eta}^{\xi\lambda\tau}}.
\label{eq:q-sec-deri}
\end{align}
In the last step, the contracted representation of the Bianchi identity
for spaces with torsion was inserted\cite{plebanski06}
\begin{align*}
R\indices{^{\alpha}_{\tau\beta\lambda}}+
R\indices{^{\alpha}_{\beta\lambda\tau}}+R\indices{^{\alpha}_{\lambda\tau\beta}}&=
-2\left(s\indices{^{\alpha}_{\tau\beta;\lambda}}+s\indices{^{\alpha}_{\beta\lambda;\tau}}+s\indices{^{\alpha}_{\lambda\tau;\beta}}\right)\\
&\quad-4\left(s\indices{^{\alpha}_{\tau\sigma}}s\indices{^{\sigma}_{\beta\lambda}}\!+
s\indices{^{\alpha}_{\beta\sigma}}s\indices{^{\sigma}_{\lambda\tau}}\!+
s\indices{^{\alpha}_{\lambda\sigma}}s\indices{^{\sigma}_{\tau\beta}}\right),
\end{align*}
hence
\begin{align}
R\indices{^{\beta}_{\tau\beta\lambda}}+
R\indices{^{\beta}_{\beta\lambda\tau}}+R\indices{^{\beta}_{\lambda\tau\beta}}&=
-2\left(s\indices{^{\beta}_{\tau\beta;\lambda}}+s\indices{^{\beta}_{\beta\lambda;\tau}}+
s\indices{^{\beta}_{\lambda\tau;\beta}}\right)\label{eq:bianci-contr}\\
&\quad-4\left(\cancel{s\indices{^{\beta}_{\tau\sigma}}s\indices{^{\sigma}_{\beta\lambda}}}\!+\!
s\indices{^{\beta}_{\beta\sigma}}s\indices{^{\sigma}_{\lambda\tau}}\!-\!
\cancel{s\indices{^{\sigma}_{\tau\beta}}s\indices{^{\beta}_{\sigma\lambda}}}\right).\nonumber
\end{align}
Equation~(\ref{eq:q-sec-deri}) can now be equated with the covariant $\lambda$-deriva\-tive
of the right-hand side of Eq.~(\ref{eq:feq-conn-coeff3}):
\begin{align*}
&\quad\onehalf R\indices{^{\tau}_{\eta\beta\lambda}}\tilde{q}\indices{_{\tau}^{\xi\lambda\beta}}-
\onehalf \tilde{q}\indices{_{\eta}^{\tau\lambda\beta}}R\indices{^{\xi}_{\tau\beta\lambda}}-
s\indices{^{\tau}_{\beta\lambda}}\,\tilde{q}\indices{_{\eta}^{\xi\lambda\beta}_{;\tau}}
+2\left(s\indices{^{\beta}_{\tau\beta;\lambda}}+\onehalf
s\indices{^{\beta}_{\lambda\tau;\beta}}+s\indices{^{\beta}_{\sigma\beta}}
s\indices{^{\sigma}_{\tau\lambda}}\right)\tilde{q}\indices{_{\eta}^{\xi\lambda\tau}}\\
&=\left(-a_{\eta}\tilde{p}^{\,\xi\lambda}-2g_{\beta\eta}\tilde{k}^{\,\beta\xi\lambda}+
s\indices{^{\,\lambda}_{\beta\alpha}}\,\tilde{q}\indices{_{\eta}^{\xi\beta\alpha}}+
2s\indices{^{\alpha}_{\beta\alpha}}\,\tilde{q}\indices{_{\eta}^{\xi\lambda\beta}}\right)_{;\lambda}.
\end{align*}
Inserting the canonical field equations~(\ref{eq:feq-vecfield}), (\ref{eq:em-tensor-den}), and~(\ref{eq:feqs-Hdyn}) yields
\begin{align*}
&\quad\tilde{q}\indices{_{\eta}^{\tau\lambda\beta}}\pfrac{\HCdtD}{\tilde{q}
\indices{_{\xi}^{\tau\beta\lambda}}}-
\pfrac{\HCdtD}{\tilde{q}\indices{_{\tau}^{\eta\beta\lambda}}}
\tilde{q}\indices{_{\tau}^{\xi\lambda\beta}}+
\bcancel{s\indices{^{\tau}_{\beta\lambda}}\,\tilde{q}\indices{_{\eta}^{\xi\beta\lambda}_{;\tau}}}
+2\left(\xcancel{s\indices{^{\beta}_{\tau\beta;\lambda}}}+\cancel{\onehalf
s\indices{^{\beta}_{\lambda\tau;\beta}}}+s\indices{^{\beta}_{\sigma\beta}}
s\indices{^{\sigma}_{\tau\lambda}}\right)\tilde{q}\indices{_{\eta}^{\xi\lambda\tau}}\\
&=a_{\eta}\left(\pfrac{\HCdtz}{a_{\xi}}-2\tilde{p}^{\,\xi\beta}s\indices{^{\,\alpha}_{\beta\alpha}}\right)-
\pfrac{\HCdtz}{\tilde{p}^{\eta\lambda}}\tilde{p}^{\,\xi\lambda}+2g_{\beta\eta}
\left(\pfrac{\HCdtz}{g_{\beta\xi}}+\pfrac{\HCdtD}{g_{\beta\xi}}-2\tilde{k}^{\,\beta\xi\tau}
s\indices{^{\,\alpha}_{\tau\alpha}}\right)-
2\pfrac{\HCdtD}{\tilde{k}\indices{^{\beta\eta\lambda}}}\tilde{k}^{\,\beta\xi\lambda}\\
&\quad\mbox{}+\cancel{s\indices{^{\,\lambda}_{\beta\alpha;\lambda}}\,\tilde{q}\indices{_{\eta}^{\xi\beta\alpha}}}+
\bcancel{s\indices{^{\,\lambda}_{\beta\alpha}}\,\tilde{q}\indices{_{\eta}^{\xi\beta\alpha}_{;\lambda}}}+
\xcancel{2s\indices{^{\alpha}_{\beta\alpha;\lambda}}\,\tilde{q}\indices{_{\eta}^{\xi\lambda\beta}}}+
2s\indices{^{\alpha}_{\beta\alpha}}\,\tilde{q}\indices{_{\eta}^{\xi\lambda\beta}_{;\lambda}}.
\end{align*}
Sorting the terms and relabeling some formal indices gives
\begin{align*}
&\quad 2\pfrac{\HCdtD}{\tilde{k}\indices{^{\beta\eta\lambda}}}\tilde{k}^{\,\beta\xi\lambda}-
2g_{\beta\eta}\pfrac{\HCdtD}{g_{\beta\xi}}+
\pfrac{\HCdtD}{\tilde{q}\indices{_{\tau}^{\eta\beta\lambda}}}\tilde{q}\indices{_{\tau}^{\xi\beta\lambda}}-
\tilde{q}\indices{_{\eta}^{\tau\beta\lambda}}\pfrac{\HCdtD}{\tilde{q}\indices{_{\xi}^{\tau\beta\lambda}}}\\
&=a_{\eta}\pfrac{\HCdtz}{a_{\xi}}-\pfrac{\HCdtz}{\tilde{p}^{\eta\lambda}}\tilde{p}^{\,\xi\lambda}+
2g_{\beta\eta}\pfrac{\HCdtz}{g_{\beta\xi}}-2B\indices{_{\eta}^{\xi}}
\end{align*}
with
\begin{equation}\label{eq:cons-proof}
B\indices{_{\eta}^{\xi}}=s\indices{^{\,\alpha}_{\lambda\alpha}}\left(a_{\eta}\tilde{p}^{\,\xi\lambda}+
2g_{\eta\beta}\tilde{k}^{\,\beta\xi\lambda}-\tilde{q}\indices{_{\eta}^{\xi\beta\tau}}s\indices{^{\lambda}_{\beta\tau}}+
\tilde{q}\indices{_{\eta}^{\xi\lambda\beta}_{;\beta}}\right).
\end{equation}
Inserting the covariant divergence of $\tilde{q}\indices{_{\eta}^{\xi\lambda\beta}}$
from Eq.~(\ref{eq:feq-conn-coeff3}) into Eq.~(\ref{eq:cons-proof}) all terms but one cancel
\begin{align*}
B\indices{_{\eta}^{\xi}}=2\tilde{q}\indices{_{\eta}^{\xi\lambda\beta}}
s\indices{^{\,\tau}_{\beta\tau}}s\indices{^{\,\alpha}_{\lambda\alpha}}=0.
\end{align*}
Yet, this term vanishes as the product of the torsion tensors is symmetric in $\beta$ and $\lambda$,
which completes the proof of the consistency equation~(\ref{eq:consistency}).
\section{Derivation of the invariant action of the U(1)$\times$Diff(M) symmetry group\label{app:u1xM}}
The sets of local coordinates referring to two coordinate charts of the spacetime manifold $M$ are denoted by $x$ and $X$.
The extension of the $\mathrm{U}(1)$ symmetry from Sect.~\ref{sec:u1} to
the additional symmetry under the diffeomorphism group $\mathrm{Diff}(M)$
is derived from the following extended generating function of type
$\FCd_{2}^{\mu}\big(\phi,\tilde{\Pi}^{*},\phi^{*},\tilde{\Pi},a,\tilde{P},g,\tilde{K},\gamma,\tilde{Q},x\big)$:
\begin{align}
\sla{\FCd_{2}^{\mu}}{x}&=\Bigg[\tilde{\Pi}^{*\,\beta}(X)\,\phi(x)\,e^{\rmi\Lambda(x)}+
\phi^{*}(x)\,\tilde{\Pi}^{\beta}(X)\,e^{-\rmi\Lambda(x)}\nonumber\\
&\quad+\tilde{P}^{\eta\beta}(X)\left(a_{\xi}(x)+
\frac{1}{q}\pfrac{\Lambda(x)}{x^{\xi}}\right)\pfrac{x^{\xi}}{X^{\eta}}\nonumber\\
&\quad+\tilde{K}^{\alpha\lambda\beta}(X)\,g_{\xi\zeta}(x)\,\pfrac{x^{\xi}}{X^{\alpha}}\pfrac{x^{\zeta}}{X^{\lambda}}\nonumber\\
&\quad+\tilde{Q}\indices{_{\eta}^{\alpha\xi\beta}}(X)
\left(\gamma\indices{^{\kappa}_{\tau\sigma}}(x)\pfrac{X^{\eta}}{x^{\kappa}}\pfrac{x^{\tau}}{X^{\alpha}}
\pfrac{x^{\sigma}}{X^{\xi}}+\pfrac{X^{\eta}}{x^{\kappa}}\ppfrac{x^{\kappa}}{X^{\alpha}}{X^{\xi}}\right)\Bigg]
\pfrac{x^{\mu}}{X^{\beta}}\left|\pfrac{x}{X}\right|^{-1}.
\label{eq:gen-u1}
\end{align}
According to the general rules for extended canonical transformations\cite{struckmeier17a},
the particular rules from the generating function~(\ref{eq:gen-u1}) follow for the scalar fields and their conjugates as
\begin{align}
\delta_{\nu}^{\mu}\Phi(X)&=\pfrac{\FCd_{2}^{\kappa}}{\tilde{\Pi}^{*\,\nu}}
\pfrac{X^{\mu}}{x^{\kappa}}\left|\pfrac{x}{X}\right|=\delta_{\nu}^{\mu}\phi(x)e^{\rmi\Lambda(x)}\label{eq:trafophi}\\
\delta_{\nu}^{\mu}\Phi^{*}(X)&=\pfrac{\FCd_{2}^{\kappa}}{\tilde{\Pi}^{\nu}}
\pfrac{X^{\mu}}{x^{\kappa}}\left|\pfrac{x}{X}\right|=\delta_{\nu}^{\mu}\phi^{*}(x)e^{-\rmi\Lambda(x)}\label{eq:trafoaphi}\\
\tilde{\pi}^{\mu}(x)&=\pfrac{\FCd_{2}^{\mu}}{\phi^{*}}=
\tilde{\Pi}^{\beta}(X)e^{-\rmi\Lambda(x)}\pfrac{x^{\mu}}{X^{\beta}}\left|\pfrac{x}{X}\right|^{-1}\label{eq:trafopi}\\
\tilde{\pi}^{*\,\mu}(x)&=\pfrac{\FCd_{2}^{\mu}}{\phi}=
\tilde{\Pi}^{*\,\beta}(X)e^{\rmi\Lambda(x)}\pfrac{x^{\mu}}{X^{\beta}}\left|\pfrac{x}{X}\right|^{-1}\label{eq:trafoapi}.
\end{align}
For the vector field $a_{\xi}$, one encounters the \emph{inhomogeneous} rule
\begin{align}
\delta_{\beta}^{\mu}A_{\alpha}(X)&=\pfrac{\FCd_{2}^{\kappa}}{\tilde{P}^{\alpha \beta}}
\pfrac{X^{\mu}}{x^{\kappa}}\left|\pfrac{x}{X}\right|=\delta_{\beta}^{\mu}\left(a_{\xi}(x)+\frac{1}{q}\pfrac{\Lambda(x)}{x^{\xi}}\right)
\pfrac{x^{\xi}}{X^{\alpha}}\label{eq:trafoa}\\
\tilde{p}^{\mu\nu}(x)&=\pfrac{\FCd_{2}^{\nu}}{a_{\mu}}=
\tilde{P}(X)^{\alpha\beta}\pfrac{x^{\mu}}{X^{\alpha}}\pfrac{x^{\nu}}{X^{\beta}}\left|\pfrac{x}{X}\right|^{-1}.\label{eq:trafop}
\end{align}
The rules for the metric $g_{\mu\nu}$ and the connection $\gamma\indices{^{\kappa}_{\tau\sigma}}$,
in conjunction with their respective conjugates, $\tilde{k}^{\alpha\lambda\beta}$ and $\tilde{q}\indices{_{\eta}^{\alpha\xi\beta}}$, are
\begin{align}
\delta_{\beta}^{\mu}G_{\alpha\lambda}&=
\pfrac{\FCd_{2}^{\kappa}}{\tilde{K}^{\alpha\lambda\beta}}\pfrac{X^{\mu}}{x^{\kappa}}\left|\pfrac{x}{X}\right|=
g_{\xi\zeta}\,\pfrac{x^{\xi}}{X^{\alpha}}\pfrac{x^{\zeta}}{X^{\lambda}}\delta_{\beta}^{\mu}\\
\tilde{k}^{\xi\zeta\mu}&=\pfrac{\FCd_{2}^{\mu}}{g_{\xi\zeta}}=
\tilde{K}^{\alpha\lambda\beta}\pfrac{x^{\xi}}{X^{\alpha}}
\pfrac{x^{\zeta}}{X^{\lambda}}\pfrac{x^{\mu}}{X^{\beta}}\left|\pfrac{x}{X}\right|^{-1}\\
\delta_{\nu}^{\mu}\Gamma\indices{^{\eta}_{\alpha\xi}}&=
\pfrac{\FCd_{2}^{\kappa}}{\tilde{Q}\indices{_{\eta}^{\alpha\xi\nu}}}
\pfrac{X^{\mu}}{x^{\kappa}}\left|\pfrac{x}{X}\right|\\
&=\delta_{\nu}^{\mu}\left(\gamma\indices{^{\kappa}_{\tau\sigma}}
\pfrac{X^{\eta}}{x^{\kappa}}\pfrac{x^{\tau}}{X^{\alpha}}\pfrac{x^{\sigma}}{X^{\xi}}+
\pfrac{X^{\eta}}{x^{\kappa}}\ppfrac{x^{\kappa}}{X^{\alpha}}{X^{\xi}}\right)\label{eq:trafogamma}\\
\tilde{q}\indices{_{k}^{\tau\sigma\mu}}&=\pfrac{\FCd_{2}^{\mu}}{\gamma\indices{^{\kappa}_{\tau\sigma}}}=
\tilde{Q}\indices{_{\eta}^{\alpha\xi\lambda}}\pfrac{X^{\eta}}{x^{\kappa}}
\pfrac{x^{\tau}}{X^{\alpha}}\pfrac{x^{\sigma}}{X^{\xi}}\pfrac{x^{\mu}}{X^{\lambda}}\left|\pfrac{x}{X}\right|^{-1}.
\end{align}
Finally, the transformation rule for the Hamiltonian follows from the \emph{explicit} dependence of the generating function,
\begin{equation}
\sla{\tilde{\HCd}^{\prime}_{0}}{X}=\left(\sla{\HCdtz}{x}+
\left.\pfrac{\FCd_{2}^{\alpha}}{x^{\alpha}}\right|_{\mathrm{expl}} \right)\left|\pfrac{x}{X}\right|.
\label{eq:Hamilton}
\end{equation}
Owing to Eq.~(A3) of Ref.\cite{struckmeier17a},
\begin{align*}
\pfrac{}{x^{\alpha}}\left(\pfrac{x^{\alpha}}{X^{\beta}}\left|\pfrac{x}{X}\right|^{-1}\right)\equiv0,
\end{align*}
the divergence of the explicitly $x$-dependent coefficients of $\FCd_{2}^{\mu}$ simplifies to
\begin{align}
&\left.\pfrac{\FCd_{2}^{\alpha}}{x^{\alpha}}\right|_{\mathrm{expl}}=
\left|\pfrac{x}{X}\right|^{-1}\Bigg\{\left(\tilde{\Pi}^{*\,\alpha}\phi\,e^{\rmi\Lambda}-
\phi^{*}\tilde{\Pi}^{\alpha}e^{-\rmi\Lambda}\right)\rmi\pfrac{\Lambda}{X^{\alpha}}\nonumber\\
&+\tilde{P}^{\eta\alpha}\left[\left(a_{\xi}+\frac{1}{q}\pfrac{\Lambda}{x^{\xi}}\right)
\ppfrac{x^{\xi}}{X^{\eta}}{X^{\alpha}}+\frac{1}{q}\ppfrac{\Lambda}{x^{\xi}}{x^{\beta}}
\xX{\xi}{\eta}\xX{\beta}{\alpha}\right]\nonumber\\
&+\tilde{K}^{\alpha\lambda\beta}g_{\xi\zeta}
\left(\ppfrac{x^{\xi}}{X^{\alpha}}{X^{\beta}}\pfrac{x^{\zeta}}{X^{\lambda}}+
\ppfrac{x^{\zeta}}{X^{\lambda}}{X^{\beta}}\pfrac{x^{\xi}}{X^{\alpha}}\right)\nonumber\\
&+\tilde{Q}\indices{_{\eta}^{\alpha\xi\beta}}\left[\gamma\indices{^{\kappa}_{\tau\sigma}}\pfrac{}{X^{\beta}}\left(
\pfrac{X^{\eta}}{x^{\kappa}}\pfrac{x^{\tau}}{X^{\alpha}}\pfrac{x^{\sigma}}{X^{\xi}}\right)+
\pfrac{}{X^{\beta}}\left(\pfrac{X^{\eta}}{x^{\kappa}}\ppfrac{x^{\kappa}}{X^{\alpha}}{X^{\xi}}\right)\right]\Bigg\}.
\label{eq:gf-p0}
\end{align}
All terms depending on the phase $\Lambda(x)$ can now be expressed
in terms of the system's fields according to the canonical transformation rules~(\ref{eq:trafophi}) to~(\ref{eq:trafoa}).
The first line of Eq.~(\ref{eq:gf-p0}) yields
\begin{align}
&\quad\left(\tilde{\Pi}^{*\,\alpha}\phi\,e^{\rmi\Lambda}-
\phi^{*}\tilde{\Pi}^{\alpha}e^{-\rmi\Lambda}\right)
\rmi\pfrac{\Lambda}{X^{\alpha}}\left|\pfrac{x}{X}\right|^{-1}\nonumber\\
&=\rmi q\left(\tilde{\Pi}^{*\,\alpha}\phi\,e^{\rmi\Lambda}-
\phi^{*}\tilde{\Pi}^{\alpha}e^{-\rmi\Lambda}\right)\left|\pfrac{x}{X}\right|^{-1}
\left(A_{\alpha}-a_{\xi}\xX{\xi}{\alpha}\right)\nonumber\\
&=\rmi q\left(\tilde{\Pi}^{*\,\alpha}A_{\alpha}\Phi-\Phi^{*}A_{\alpha}\tilde{\Pi}^{\alpha}\right)\left|\pfrac{x}{X}\right|^{-1}-
\rmi q\left(\tilde{\pi}^{*\,\alpha}a_{\alpha}\phi-\phi^{*}a_{\alpha}\tilde{\pi}^{\alpha}\right)
\label{eq:gf-p1}
\end{align}
Hence all $\Lambda$-dependent terms are eliminated and replaced in a symmetric way by the fields
of the original and the transformed system.
The second line of Eq.~(\ref{eq:gf-p0}) is treated similarly by inserting Eqs.~(\ref{eq:trafoa}) and~(\ref{eq:trafop})
\begin{align}
&\quad\tilde{P}^{\eta\beta}\left[\left(a_{\xi}+\frac{1}{q}\pfrac{\Lambda}{x^{\xi}}\right)
\ppfrac{x^{\xi}}{X^{\eta}}{X^{\beta}}+\frac{1}{q}\ppfrac{\Lambda}{x^{\xi}}{x^{\alpha}}
\xX{\xi}{\eta}\xX{\alpha}{\beta}\right]\left|\pfrac{x}{X}\right|^{-1}\nonumber\\
&=\left[\tilde{P}^{\eta\beta}A_{\alpha}\Xx{\alpha}{\xi}\ppfrac{x^{\xi}}{X^{\eta}}{X^{\beta}}
+\tilde{P}^{(\eta\beta)}\xX{\xi}{\eta}\xX{\alpha}{\beta}\pfrac{}{x^{\alpha}}
\left(A_{\tau}\Xx{\tau}{\xi}-a_{\xi}\right)\right]\left|\pfrac{x}{X}\right|^{-1}\nonumber\\
&=\frac{1}{2}\tilde{P}^{\eta\beta}\left(\pfrac{A_{\eta}}{X^{\beta}}+
\pfrac{A_{\beta}}{X^{\eta}}\right)\left|\pfrac{x}{X}\right|^{-1}-
\frac{1}{2}\tilde{p}^{\eta\beta}\left(\pfrac{a_{\eta}}{x^{\beta}}+\pfrac{a_{\beta}}{x^{\eta}}\right)\nonumber\\
&\qquad+\tilde{P}^{\eta\beta}A_{\alpha}\left|\pfrac{x}{X}\right|^{-1}\underbrace{\left(
\Xx{\alpha}{\xi}\ppfrac{x^{\xi}}{X^{\eta}}{X^{\beta}}+\ppfrac{X^{\alpha}}{x^{\xi}}{x^{\tau}}
\xX{\xi}{\eta}\xX{\tau}{\beta}\right)}_{\equiv0}.
\label{eq:gf-p2}
\end{align}
The last line of Eq.~(\ref{eq:gf-p2}) vanishes by virtue of the identity
\begin{align*}
\pfrac{}{X^{\beta}}\left(\pfrac{X^{\alpha}}{x^{\xi}}\pfrac{x^{\xi}}{X^{\eta}}\right)\equiv
\pfrac{}{X^{\beta}}\left(\delta_{\eta}^{\alpha}\right)\equiv0,
\end{align*}
and thus
\begin{equation}\label{eq:Xinv}
\pfrac{X^{\alpha}}{x^{\xi}}\ppfrac{x^{\xi}}{X^{\eta}}{X^{\beta}}\equiv
-\ppfrac{X^{\alpha}}{x^{\xi}}{x^{\tau}}\pfrac{x^{\xi}}{X^{\eta}}\pfrac{x^{\tau}}{X^{\beta}}.
\end{equation}
Hence, the coupling of the term $\tilde{p}^{\eta\beta}a_{\alpha}$ to the dynamical spacetime geometry is
exactly compensated due to the inhomogeneous transformation rule~(\ref{eq:trafoa}), which
is the $\mathrm{U}(1)$ symmetry transformation rule for the gauge field $a_{\mu}$.
This is confirmed by setting $\Lambda\equiv0$ in Eq.~(\ref{eq:gen-u1}) and in the subsequent
transformation rules~(\ref{eq:trafophi}) to~(\ref{eq:gf-p0})---which amounts to eliminating
the $\mathrm{U}(1)$ symmetry transformation.
In this case, the connection coefficients pertaining to the spacetime manifold enter
as gauge quantities, which converts the partial derivative of the gauge field $a_{\mu}$
into a covariant derivative and thus renders the action integral generally covariant.

For the case of a system with additional $\mathrm{U}(1)$ symmetry, hence $\Lambda\not\equiv0$,
Eq.~(\ref{eq:gf-p0}) yields inserting Eqs.~(\ref{eq:gf-p1}) and~(\ref{eq:gf-p2})
\begin{align}
&\left.\pfrac{\FCd_{2}^{\alpha}}{x^{\alpha}}\right|_{\mathrm{expl}}\!=
\rmi q\left(\tilde{\Pi}^{*\,\alpha}\Phi-\Phi^{*}\tilde{\Pi}^{\alpha}\right)A_{\alpha}\!\left|\pfrac{x}{X}\right|^{-1}\!\!\!-
\rmi q\left(\tilde{\pi}^{*\,\alpha}\phi-\phi^{*}\tilde{\pi}^{\alpha}\right)a_{\alpha}\nonumber\\
&+\onehalf\tilde{P}^{\alpha\beta}\left(\pfrac{A_{\alpha}}{X^{\beta}}+
\pfrac{A_{\beta}}{X^{\alpha}}\right)\left|\pfrac{x}{X}\right|^{-1}-\onehalf\tilde{p}^{\alpha\beta}
\left(\pfrac{a_{\alpha}}{x^{\beta}}+\pfrac{a_{\beta}}{x^{\alpha}}\right)\nonumber\\
&+\left(\tilde{K}^{\alpha\lambda\beta}G_{\xi\lambda}+
\tilde{K}^{\lambda\alpha\beta}G_{\lambda\xi}\right)\Gamma\indices{^{\xi}_{\alpha\beta}}\left|\pfrac{x}{X}\right|^{-1}
\!\!\!-\left(\tilde{k}^{\alpha\lambda\beta}g_{\xi\lambda}+
\tilde{k}^{\lambda\alpha\beta}g_{\lambda\xi}\right)\gamma\indices{^{\xi}_{\alpha\beta}}\nonumber\\
&+\onehalf\tilde{Q}\indices{_{\eta}^{\alpha\xi\beta}}\left(
\pfrac{\Gamma\indices{^{\eta}_{\alpha\xi}}}{X^{\beta}}+\pfrac{\Gamma\indices{^{\eta}_{\alpha\beta}}}{X^{\xi}}+
\Gamma\indices{^{k}_{\alpha\beta}}\Gamma\indices{^{\eta}_{k\xi}}-
\Gamma\indices{^{k}_{\alpha\xi}}\Gamma\indices{^{\eta}_{k\beta}}\right)\left|\pfrac{x}{X}\right|^{-1}\nonumber\\
&-\onehalf\tilde{q}\indices{_{\eta}^{\alpha\xi\beta}}\left(
\pfrac{\gamma\indices{^{\eta}_{\alpha\xi}}}{x^{\beta}}+\pfrac{\gamma\indices{^{\eta}_{\alpha\beta}}}{x^{\xi}}+
\gamma\indices{^{k}_{\alpha\beta}}\gamma\indices{^{\eta}_{k\xi}}-
\gamma\indices{^{k}_{\alpha\xi}}\gamma\indices{^{\eta}_{k\beta}}\right).
\label{eq:gf-fin}
\end{align}
The terms in Eq.~(\ref{eq:gf-p0}) proportional to $\tilde{K}^{\alpha\lambda\beta}$ and
$\tilde{Q}\indices{_{\eta}^{\alpha\xi\beta}}$ were already discussed in Ref.\cite{struckmeier17a}.
We observe that the divergence of the spacetime dependent coefficients of $\FCd_{2}^{\mu}$
is expressed symmetrically in the original and the transformed fields.
The particular Hamiltonian
\begin{align}
\tilde{\HCd}_{3}&=\HCdtz+
\rmi q\left(\tilde{\pi}^{*\,\alpha}\phi-\phi^{*}\tilde{\pi}^{\alpha}\right)a_{\alpha}+\onehalf\tilde{p}^{\alpha\beta}
\left(\pfrac{a_{\alpha}}{x^{\beta}}+\pfrac{a_{\beta}}{x^{\alpha}}\right)\nonumber\\
&\quad+\left(\tilde{k}^{\alpha\lambda\beta}g_{\xi\lambda}+
\tilde{k}^{\lambda\alpha\beta}g_{\lambda\xi}\right)\gamma\indices{^{\xi}_{\alpha\beta}}
+\onehalf\tilde{q}\indices{_{\eta}^{\alpha\xi\beta}}\left(
\pfrac{\gamma\indices{^{\eta}_{\alpha\xi}}}{x^{\beta}}+\pfrac{\gamma\indices{^{\eta}_{\alpha\beta}}}{x^{\xi}}+
\gamma\indices{^{\kappa}_{\alpha\beta}}\gamma\indices{^{\eta}_{\kappa\xi}}-
\gamma\indices{^{\kappa}_{\alpha\xi}}\gamma\indices{^{\eta}_{\kappa\beta}}\right)\label{eq:Hamilton1}
\end{align}
is thus converted by Eq.~(\ref{eq:gf-fin}) according to the general transformation rule~(\ref{eq:Hamilton})
into a Hamiltonian $\tilde{\HCd}_{3}^{\prime}$ of the same form.
Combining the initial action integral~(\ref{eq:action-integral2}) with the ``gauged''
Hamiltonian~(\ref{eq:Hamilton1}), the form-invariant action integral follows as
\begin{align}
&S=\int\left[\tilde{\pi}^{*\,\alpha}\left(\pfrac{\phi}{x^{\alpha}}-\rmi q\,a_{\alpha}\phi\right)+
\left(\pfrac{\phi^{*}}{x^{\alpha}}+\rmi q\,\phi^{*}a_{\alpha}\right)\tilde{\pi}^{\alpha}\right.\nonumber\\
&\mbox{}+\onehalf\tilde{p}^{\,\alpha\beta}\left(\pfrac{a_{\alpha}}{x^{\beta}}-
\pfrac{a_{\beta}}{x^{\alpha}}\right)+\tilde{k}^{\,\alpha\lambda\beta}\left(\pfrac{g_{\alpha\lambda}}{x^{\,\beta}}-
g_{\xi\lambda}\gamma\indices{^{\xi}_{\alpha\beta}}-g_{\alpha\xi}\gamma\indices{^{\xi}_{\lambda\beta}}\right)\nonumber\\
&\mbox{}+\onehalf\tilde{q}\indices{_{\eta}^{\alpha\xi\beta}}\left(
\pfrac{\gamma\indices{^{\eta}_{\alpha\xi}}}{x^{\beta}}-\pfrac{\gamma\indices{^{\eta}_{\alpha\beta}}}{x^{\xi}}-
\gamma\indices{^{\kappa}_{\alpha\beta}}\gamma\indices{^{\eta}_{\kappa\xi}}+
\gamma\indices{^{\kappa}_{\alpha\xi}}\gamma\indices{^{\eta}_{\kappa\beta}}\right)-\HCdtz\Bigg]\dd^{4}x.
\label{eq:action3}
\end{align}
As both integrands are world scalar densities, the action integrals are form-invariant
under both the $\mathrm{U}(1)$ symmetry group \emph{and} under the diffeomorphism group.
In particular, the partial derivatives of the complex scalar fields are amended according to Eq.~(\ref{eq:Hamilton1})
to yields the \emph{gauge covariant} derivatives and hence implement the \emph{minimum coupling principle}---which
is not postulated here but emerges from the canonical transformation formalism.
Furthermore, the vector field $a_{\mu}$ does not couple \emph{directly} to the
spacetime geometry as the respective term cancels in Eq.~(\ref{eq:gf-p2}).
The covariant derivative of $a_{\mu}$ in the generally covariant action integral~(\ref{eq:action-integral4})
is replaced due to the $\mathrm{U}(1)$ symmetry by the curl of $a_{\mu}$ in Eq.~(\ref{eq:action3}), which already has tensor property.
The coupling thus occurs only via the related energy-momentum tensor and not via the connections coefficients.
This statement also holds for the $\mathrm{SU}(N)$ symmetry group.

\end{document}